\documentclass[a4paper,fleqn, times]{cas-dc}

\usepackage[numbers]{natbib}
\usepackage{subfigure}
\usepackage{color}
\usepackage{multirow}
\usepackage{graphicx}
\usepackage{chngcntr}
\usepackage{hyperref}
\usepackage[switch]{lineno}
\usepackage{amssymb}
\usepackage{tabularx}              
\usepackage{amsmath}
\usepackage{array}  
\usepackage[most]{tcolorbox}
\usepackage{xcolor}
\usepackage{longtable} %
\usepackage{xurl} 

\begin{document}
\let\WriteBookmarks\relax
\def\floatpagepagefraction{1}
\def\textpagefraction{.001}

\shorttitle{A Grey Literature Review of AI-Native Applications}

\shortauthors{Cao et al.}   

\title [mode = title]{A Grey Literature Review of AI-Native Applications}

\author[1]{Lingli Cao}
\ead{cll@smail.nju.edu.cn}

\author[1]{Shanshan Li}
\cormark[1]
\ead{lss@nju.edu.cn}

\author[1]{Ying Fan}
\ead{3470408932@qq.com}

\author[1]{Danyang Li}
\ead{522024320080@smail.nju.edu.cn}

\author[2]{Chenxing Zhong}
\ead{chenxingzhong@njust.edu.cn}
            
\affiliation[1]{organization={State Key Laboratory of Novel Software Technology, Software Institute, Nanjing University},
            city={Nanjing},
            country={China}}

\affiliation[2]{organization={Nanjing University of Science and Technology}, 
            city={Nanjing},
            country={China}}

\cortext[1]{Corresponding author}

\begin{abstract}
\noindent \textbf{Context:} 
AI-native applications leverage generative AI as the core engine driving architecture. Despite substantial investment and rapid practical adoption, theoretical foundations remain underdeveloped, with interpretations varying across the industry and academic research still nascent. This gap creates inefficiencies and risks that hinder large-scale adoption.

\noindent \textbf{Objective:}
To facilitate the successful adoption of AI-native applications, it is necessary to establish a comprehensive architectural-level understanding. This study seeks to identify the defining characteristics, key quality attributes, and technical implementation elements of AI-native applications, as well as clarify the opportunities and challenges they present.

\noindent \textbf{Method:}
At present, discourse on AI-native applications is predominantly shaped by practitioner-driven open-source project practices and scattered technical blog posts. Thus, We conducted a grey literature review (GLR), integrating conceptual perspectives retrieved from targeted Google and Bing searches with practical insights derived from leading open-source projects on GitHub. A structured protocol encompassing source selection, quality assessment, and thematic analysis was applied to synthesize findings across heterogeneous sources. 

\noindent \textbf{Results:}
We finally identified 106 GL sources based on the selection criteria. The analysis reveals that AI-native applications are distinguished by two core pillars: the central role of AI as the system’s intelligence paradigm and its inherent generative and multimodal capabilities. Critical quality attributes include traditional QAs such as maintainability and scalability, and AI-specific QAs such as adaptability and observability. The technical implementation elements include core model technologies, agentic systems, RAG, among others. It is noteworthy that more than 60\% of GL originates from open-source project documentation, empirically highlighting that AI-native applications are largely practice-driven, while their underlying theoretical foundations remain underdeveloped.

\noindent \textbf{Conclusion:}
This study is the first to propose a dual-layered engineering blueprint for AI-native applications. The blueprint provides actionable design guidelines and technical recommendations for practitioners, while also laying a foundation for formalizing AI-native architectural patterns and advancing future research in software engineering.

\end{abstract}


\begin{keywords} 
AI-native Applications\sep 
Generative AI\sep 
Quality Attributes\sep 
Grey Literature Review
\end{keywords}

\maketitle

\section{Introduction}
\label{sec1}
In recent years, generative AI technologies \cite{feuerriegel2024generative}, exemplified by large language models (LLMs), have achieved groundbreaking advances, fundamentally reshaping the boundaries of software development and human–computer interaction \cite{fui2023generative}. Beyond the AI-enabled paradigm, which incorporates AI technologies as auxiliary features into existing applications, the new paradigm, AI-native applications, is rapidly emerging as a transformative force in the software industry \cite{hassan2024towards}. It no longer views AI as an auxiliary feature but positions generative AI technologies as the intelligent engines that drive its core logic, user interaction, and overall system architecture, with representative examples such as ChatGPT\footnote{https://chatgpt.com/} and Midjourney\footnote{https://www.midjourney.com/home}. These applications are projected to play an increasingly central role in enterprise software, with significant market expectations and rapid adoption trends reported by industry leaders \cite{sapphire2025, hymel2024ai}. For example, an industry report by the prominent venture capital firm Sapphire Ventures notes that AI-native applications are receiving increasing attention in enterprise software development, and that their strategic design and technological practices are gradually becoming key considerations for both investment and development \cite{sapphire2025}. It is estimated that in 2025, at least 47 AI-native applications will emerge, each generating an annual recurring revenue of over 50 million USD. Similar to how the cloud-native paradigm reshaped software deployment and delivery, AI-native applications are expected to profoundly transform the ways in which software is designed, developed, and evolved.

Despite the growing prominence of AI-native applications, their theoretical foundations remain significantly underdeveloped. Although the term 'AI-native application' is widely used, its definition remains unclear, much like many concepts in the AI domain, and it is still evolving \cite{sapphire2025,alibabacloud20251}. For example, in terms of conceptual definitions, the well-known telecommunications company Ericsson states that “An AI native implementation leverages a data-driven and knowledge-based ecosystem, where data/knowledge is consumed and produced to realize new AI-based functionality or augment and replace static, rule-based mechanisms with learning and adaptive AI when needed.” \cite{ericsson_ai_native}.
The prominent venture capital firm Sapphire Ventures highlights that “AI-native applications are built on foundational AI capabilities, like learning from large datasets, understanding context, or generating novel outputs. AI-native applications deliver outcomes that break traditional constraints of speed, scale and cost, enabling entirely new possibilities...” \cite{sapphire2025}. Meanwhile, some technology visionaries in the mobile application domain argue that “AI-native apps are built with architectures that fully incorporate machine learning and AI models as fundamental components.” \cite{pluralsight2025}. Industry perceptions of AI-native applications remain fragmented. 

Meanwhile, academic research on AI-native applications remains in early stages, with relatively limited literature. Existing studies focus more on generative AI technologies themselves and on the AI-enabled paradigm, such as Reinforcement Learning from Human Feedback (RLHF) framework \cite{zheng2023secrets}, efficient fine-tuning \cite{lin2024data}, automated code reviews \cite{yu2024fine}, and architecture consistency checking \cite{cao2025enhancing}, rather than on AI-native applications as a distinct paradigm. 

This gap contributes to piecemeal exploration, inefficient resource utilization, and substantial trial-and-error costs, introducing unpredictable quality risks and ultimately constituting a key bottleneck to large-scale adoption. For example, conceptual discussions in industry white papers and strategic analyses often frame the AI-native applications paradigm in abstract business or visionary terms; the actual development, predominantly visible in open-source projects, focuses on exploring concrete implementation pathways independently through practice. This misalignment between perception and practice has resulted in a lack of systematic guidance and difficulties in knowledge consolidation. To facilitate the successful adoption of AI-native applications, it is necessary to establish a systematic architectural-level understanding. Similar to the evolution of other architectural paradigms in software engineering, during the early stages of a paradigm, it is necessary to clarify its core characteristics, key quality attributes, and technical implementation elements at the architectural level, thereby laying a theoretical foundation for its maturation. For example, following the introduction of the microservices paradigm, researchers led by Flower~\cite{martinfowler2014} systematically analyzed the core characteristics, key quality attributes, and technical aspects of microservices architecture~\cite{LI2021106449, 8703917, ZHOU2023111521}, effectively promoting its successful practice and widespread adoption.
 
This study is to provide a comprehensive synthesis of AI-native applications, focusing on their core characteristics, key quality attributes, technical implementation elements, and emerging opportunities and challenges. This study addresses the above gap by conducting a Grey Literature Review (GLR) \cite{garousi2019guidelines}, drawing on conceptual perspectives from industry reports and technical blogs as well as practical evidence from open-source projects on GitHub. Traditional Systematic Literature Review (SLR)  \cite{keele2007guidelines,KITCHENHAM20097} and Systematic Mapping Study (SMS) \cite{petersen2008systematic,PETERSEN20151} are not adopted in this study. The reason is that SLR and SMS primarily rely on peer-reviewed white literature, while in the emerging domain of AI-native applications, academic research is still at an early stage and remains limited in scope and volume. Widely recognized experts and a stable conceptual framework in this domain have yet to emerge. Other methods, such as the Delphi technique \cite{Delphi2005}, which rely on expert consensus, are less suitable in this context and may produce overly subjective and speculative results. Therefore, GLR represents a more appropriate and effective research method at the present stage, as it provides practitioners’contemporary perspectives on important topics in both practice and research, and presents empirical data derived from practitioners’real-world software development activities \cite{garousi2020benefitting,https://doi.org/10.1002/smr.2197}.

\begin{itemize}
    \item This study reviews 106 GL sources and identifies the defining characteristics and conceptual boundaries of AI-native applications.
    \item This study synthesizes the most emphasized quality attributes and their associated challenges.
    \item This study maps the technical elements supporting the development of AI-native applications.
    \item This study proposes the first dual-layered engineering blueprint for AI-native applications offering actionable guidelines for practitioners and a foundation for future research.
\end{itemize}

The remainder of this study is organized as follows. Section~\ref{sec2} reviews related work. Section~\ref{sec3} details the research methodology, including the grey literature search strategy, data selection criteria, and analysis methods. Section~\ref{sec4} presents the research findings. Section~\ref{sec5} provides an in-depth discussion of the results, and identifies potential threats to the validity of this study. Finally, Section~\ref{sec7} concludes the paper and outlines possible directions for future work.

\section{Related Work} 
\label{sec2}
\subsection{Industry perspectives}
\label{sub:industry}
AI-native applications are rapidly emerging in the industry, attracting unprecedented levels of technological investment and market attention. The prominent venture capital firm Sapphire Ventures highlighted that in 2024, investment in generative AI-native applications surged, with their proportion of total generative AI funding rising markedly compared to the previous two years \cite{sapphire2025}. According to Accel’s 2025 Globalscape report, AI models and AI-native applications are driving record funding in 2025, with total investment projected to reach \$184 billion, nearly an 80\% increase compared to the previous year \cite{accel2025}. In 2025, Gartner’s research report further emphasized that AI-native software engineering is transforming traditional development processes \cite{gartner2025}. 

However, despite the thriving AI-native practices, organizations have yet to reach a consensus on the core conceptual framework of AI-native applications, leading to understandings that are partially overlapping but largely independent. For example, the venture capital firm Andreessen Horowitz highlights that all applications leveraging AI-native workflows share the essential characteristic of translating cutting-edge models into accessible and efficient user interfaces \cite{a16z2024}. Sapphire Ventures considers AI-native applications to typically involve some degree of proprietary AI technology, such as model fine-tuning or orchestration, rather than relying entirely on off-the-shelf capabilities, emphasizing how AI technologies generate novel system outputs and enable continuous optimization \cite{sapphire2025}. In contrast, the telecommunications equipment and solutions provider Ericsson offers a definition that places greater emphasis on engineering and system architecture. They argue that AI-native is a concept of intrinsically trustworthy AI capabilities, in which AI is a natural and integral part of functionalities across design, deployment, operation, and maintenance, emphasizing both the nativeness of AI and the system's sustainable and reliable capabilities \cite{ericsson_ai_native}. Alibaba Cloud, a leading provider of cloud computing and infrastructure services, considers AI-native applications as designed to achieve enhanced model efficiency and effectiveness, with a strong emphasis on observability \cite{Alibaba2024}. Frontline technical experts and developer communities, on the other hand, place greater emphasis on concrete design patterns. For instance, Martin Fowler has systematically organized various application patterns of generative AI, such as Retrieval-Augmented Generation (RAG), providing developers with several well-established low-level design patterns \cite{fowler2024genai}.Such a fragmented understanding of AI-native applications leads to piecemeal exploration, inefficient resource utilization, and substantial trial-and-error costs, introducing unpredictable quality risks and ultimately constituting a key bottleneck to large-scale adoption. According to RAND Corporation’s 2024 research, AI projects exhibit a failure rate of up to 80\% \cite{rand2024}. At the same time, IDC forecasts that global investment in AI will reach \$630 billion by 2028 \cite{IDC2024}. This suggests that such a high failure rate could lead to hundreds of billions of dollars in wasted investment, productivity losses, and missed opportunities \cite{Ryan12025}. 

To facilitate the successful adoption of AI-native applications, it is necessary to establish a comprehensive architectural-level understanding. Similar to the evolution of other paradigms in software engineering, such as the microservices paradigm, researchers have systematically analyzed and synthesized core characteristics, key quality attributes, and technical implementation elements~\cite{martinfowler2014,7796008, 7030212, nadareishvili2016microservice, LI2021106449, 8703917, ZHOU2023111521}, gradually forming an industry-recognized microservices architectural paradigm that has effectively promoted its successful practice and widespread adoption. In view of this, it is imperative to conduct systematic research on AI-native applications to identify their core characteristics, key quality attributes, and technical implementation elements, thereby providing a solid theoretical foundation for AI-native application practice and accelerating their engineering maturation process. 

\subsection{Academic perspectives}
\label{sub:academia}
Currently, research on the AI-native applications paradigm is still in its infancy, with existing studies primarily focusing on generative AI technologies themselves and their enabling application scenarios. Much of this body of research seeks to advance generative AI technologies, including enhancing domain-specific knowledge in LLMs through fine-tuning~\cite{lin2024data,han2024parameterefficientfinetuninglargemodels}, precisely controlling LLMs outputs via prompt engineering~\cite{liu2022design,white2023prompt,santu2023teler}, aligning LLMs behavior through Reinforcement Learning from Human Feedback (RLHF)~\cite{zheng2023secrets,ziegler2019fine}, and improving performance on downstream tasks by leveraging retrieval-augmented generation (RAG)~\cite{ren2025investigating,arslan2024survey} and LLMs-based agent~\cite{li2024survey,talebirad2023multi} approaches. Building on this foundation, researchers have further investigated the feasibility of AI-enabled paradigm. For example, generative AI techniques have been applied to various stages of the software development lifecycle, such as software architecture design ~\cite{10592789,10592785,diaz2024helping}, code generation \cite{jiang2024survey,10.1145/3770084,espejel2024low}, and software testing \cite{10440574,10.1145/3650212.3680388,10.1145/3696630.3734199}, in order to support developers and improve productivity. In addition, some studies have explored the quality attributes, technical challenges, key concerns and design principles of AI-enabled applications ~\cite{scaramuzza2024accountability,martinez2022software,felderer2021quality,indykov2025architectural,gozalo2023survey,fui2023generative,weisz2024design}. Such applications generally follow traditional software architecture design principles, treating AI as an integrated module to enhance product functionality.

This stands in stark contrast to the AI-native paradigm. AI-native applications go far beyond the mere integration of AI; instead, they represent software systems that are fundamentally architected around AI-driven capabilities at their core. In this paradigm, AI models serve not as peripheral features but as the central engines that drive application logic, user experience, and even the underlying backend architecture. At present, only a limited number of forward-looking studies have called for the shift toward AI-native software engineering (SE 3.0)~\cite{hassan2024towards}, along with a corresponding software development lifecycle~\cite{hymel2024ai}. However, an architecture-level understanding of AI-native applications remains notably lacking in existing literature. Specifically, questions about the core characteristics, key quality attributes, and technical implementation elements of AI-native applications remain underexplored. To address this gap, this study extends prior work by conducting a grey literature review that synthesizes industry knowledge and practical experiences, bridging the disconnect between fragmented industry definitions and the absence of academic frameworks.

\section{Methodology} 
\label{sec3}
This study aimed to overcome the fragmented understanding of AI-native applications by conducting a Grey Literature Review
(GLR)~\cite{garousi2020benefitting} that synthesizes industry knowledge and practical experiences. It sought to construct a comprehensive architectural perspective of AI-native applications, clarify their core characteristics, key quality attributes, and technical implementation elements, and provide a solid theoretical foundation for AI-native application practice and accelerating their engineering maturation process.

\subsection{Study design}
This study followed the guidelines for conducting GLR in software engineering~\cite{garousi2019guidelines}. The review was initiated in early 2025 and involved a multidisciplinary team comprise one PhD student, two master students, and two supervisors with extensive experience in empirical research. Specifically, the PhD student was responsible for developing the review protocol, defining the research questions, and designing the data selection and extraction schemes. Subsequently, the two master students joined the remaining review process. The entire review was guided by the two supervisors, who also performed random cross-checks of the students’ results in the literature selection and extraction stages. The review team held regular meetings to check the review process and to discuss disagreements or emerging issues regarding the search strategy, inclusion and exclusion criteria, data extraction, and data synthesis methods, etc. until consensus was reached.


\subsection{Research questions}
\label{RQs}
This study was structured around four research questions. The first three formed the core content of a comprehensive architectural perspective on AI-native applications together. The fourth, in turn, examined the opportunities and challenges brought by AI-native applications from a macro perspective, highlighting the practical significance of the constructed comprehensive understanding. Specifically, it aimed to address the following central questions:

\textbf{RQ1: What are the key characteristics of AI-native applications?} This question sought to examine and analyze the fundamental characteristics of AI-native applications, identify points of consensus and divergence across existing studies and practices, and clarify both the essential and optional elements, thereby delineating the conceptual boundaries of AI-native applications.

\textbf{RQ2: Which quality attributes are most prominently considered during the design of AI-native applications?} This question sought to identify the quality attributes emphasized in AI-native applications and the challenges associated with them.

\textbf{RQ3: What are the technical elements used in the development of AI-native applications? }This question aimed to map the technical implementation elements of AI-native applications and examine the extent to which these technology choices can address the quality attribute challenges identified in RQ2.

\textbf{RQ4: What opportunities and challenges do AI-native applications present?} This question sought to uncover the potential value of the emerging AI-native paradigm and to identify the key challenges faced during its implementation, thereby providing a foundation for future academic research and industrial practice.

\subsection{Search process}
This study selected Google\footnote{https://www.google.com/}, Bing\footnote{https://www.bing.com/}, and GitHub\footnote{https://www.github.com/} as the primary data sources. This study analyzed industry leaders’ conceptual understanding of AI-native applications using Google and Bing, while simultaneously examining practitioners’ practical insights through major open-source projects on GitHub. As the most widely used search engines, Google and Bing can efficiently cover various grey literature resources dispersed across the web \cite{esposito2025generative, bogner2021industry}, such as blogs, white papers, and technical reports. As the world's largest open-source community, GitHub hosts a wealth of project documentation, issue reports, and practical experiences, and is also regarded as an important source of grey literature \cite{zhang2020evidence}.



In designing the search strings, we followed a multi-source input process. The search strings were developed based on three dimensions: 1) industry insights, derived from our prior interview experiences with industry leaders such as Alibaba Cloud; 2) research objectives aligned with our research questions; and 3) expert feedback from supervisors. 

Note that to capture relevant GL sources as comprehensively and accurately as possible, we did not include generative AI technologies in our search strings. Although AI-native applications were closely associated with generative AI in the current context, this study aimed to objectively analyze the paradigm characteristics of AI-native applications. Therefore, we restricted our search to literature explicitly labeled as involving AI-native applications. The final search strings employed in this study was as follows. The first set of terms encompassed vocabulary associated with AI-native concepts, whereas the second set narrowed the search to applications. Within each set, terms were combined using the OR operator, while the AND operator was used to link terms across different sets.

\begin{tcolorbox}[colback=gray!20, colframe=gray!20]
("AI native" OR AI-native)\\
AND\\
(application OR system OR software)
\end{tcolorbox}

The search was restricted to the period between December, 2022 (following the release of ChatGPT\footnote{https://openai.com/index/chatgpt/}) and May, 2025. To minimize potential biases arising from dynamic caching, session tracking, and personalized algorithms, we carried out six independent search sessions on Google and Bing to ensure the comprehensiveness and robustness of the retrieved results. 
For each query, we relied on the search engine’s page ranking algorithm~\cite{langville2011google} and analyzed only a manageable number of highly relevant results. We adopted an effort bounded criteria~\cite{garousi2019guidelines}, i.e., including only the top N search engine hits, as the stopping criterion for this GLR search. Specifically, we collected the top 10 pages of results from Google and Bing with 10 URLs per page, following prior GLR in the software engineering domain~\cite{garousi2019guidelines, bogner2021industry}. This resulted in a total of 1,200 URLs (6*100*2). Beyond the tenth results page, no additional relevant or substantive sources were identified. For GitHub, we employed a programmatic retrieval approach, querying data through its official API to ensure the process was reproducible. Considering the varying quality and impact of open-source projects, and referring to existing empirical GitHub-based project selection criteria~\cite{10.1145/3377811.3380412,HONFI2020106319,Braganholo2019}, we filtered for projects with at least 1,000 stars, over 50 forks, and one active commit within the past year. These quantitative metrics acted as effective proxies for community authority, developer engagement, and technological timeliness, allowing us to precisely target core AI-native application practices that were both widely endorsed by the community and actively evolving.

\begin{figure}
\centering
\includegraphics[width=1.0\linewidth]{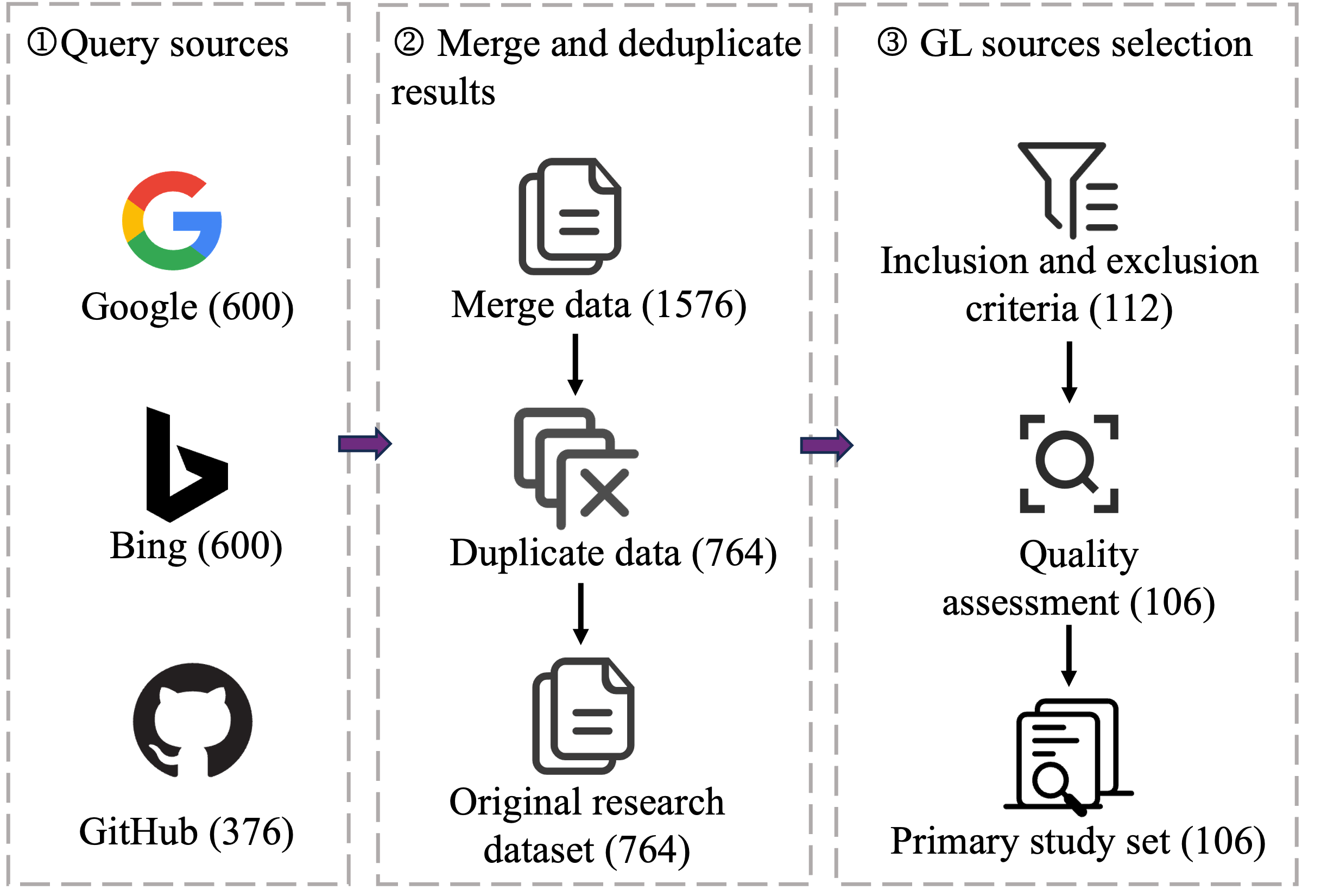}
\caption{GL sources identification process.}
\label{process}
\end{figure}

The overall retrieval process was shown in Figure~\ref{process}. After collecting search results from Google, Bing, and GitHub, we compiled them into a unified dataset of 1,576 entries. To ensure consistency during merging, all entries were standardized by URL and title. Cross-platform duplicates were identified and removed using URL comparisons. After removing duplicates, the original research dataset comprised 764 GL sources. We subsequently screened these GL sources by examining their actual content in accordance with the inclusion and exclusion criteria (see Subsection \ref{study selection}), excluded 652 entries. Then, we performed a quality assessment on the remaining 112 GL sources (see Subsection \ref{Quality assessment}). Ultimately, the GLR process yielded 106 GL sources.

\subsection{Source selection}
\label{study selection}
The inclusion and exclusion criteria applied in this study were summarized in Table~\ref{tab2}. The original research datasets were assessed against these criteria, on the basis of which selection decisions were made. The grey literature included primarily focused on AI-native applications, covering, though not limited to, their underlying concepts, defining characteristics, relevant quality attributes, development approaches, advantages, and challenges. Grey literature that was unrelated to AI-native applications, or that merely made a passing reference to them, was excluded. The source selection process was carried out with independent double-checking by graduate students. Any disagreements were resolved through meetings and further discussion with their supervisor.

\begin{table*}[!htbp] 
\centering
\caption{Selection criteria.}
\label{tab2}
\begin{tabularx}{\textwidth}{lX}
\hline
\multicolumn{2}{l}{\color[HTML]{333333}\textbf{Inclusion and exclusion criteria}} \\ \hline
I1 & The source primarily focuses on AI-native applications. \\
I2 & The source provides descriptions of the characteristics, quality attributes, or technical implementation elements of AI-native applications. \\
I3 & The source discusses the advantages or challenges associated with AI-native applications. \\
I4 & The source is not formally peer reviewed nor formally published, such as a blog, post, white paper, project documentation or preprint. \\
E1 & The source is not written in English. \\
E2 & The primary purpose of the source is commercial promotion, recruitment, or training. \\
E3 & The source does not specifically focus on AI-native applications. \\ \hline
\end{tabularx}
\end{table*}

\subsection{Source quality assessment}
\label{Quality assessment}
While researchers in software engineering are increasingly drawing on knowledge from grey literature, these sources lack peer review and exhibit considerable variability in quality \cite{garousi2020benefitting}. To assess the overall quality and credibility of each selected grey literature source, we followed the grey literature quality assessment criteria proposed by Garousi et al \cite{garousi2019guidelines} (as shown in Appendix~\ref{app1}). It includes criteria such as the information provider’s authority, methodology, objectivity, publication date, relevance to other sources, originality, impact, and publication type, each accompanied by corresponding evaluation questions. Based on the type of research data, the two authors selected appropriate quality criteria and evaluated them using either a binary Likert scale (1$=$yes, 0$=$no) or a three-point Likert scale (1$=$yes, 0$=$no, 0.5$=$neutral), according to the specific criterion. In cases of disagreement, a third author convened a meeting to determine the outcome. Finally, for each GL source, the mean score across all evaluation items was calculated as its quality score. GL sources with a quality score below 0.5 were excluded.

\subsection{Data extraction}
To systematically extract information from the selected grey literature and address the research questions of this study, we designed and employed a structured data extraction form, as presented in Table~\ref{tab3}. The first four metadata items captured the basic characteristics of the grey literature, while the remaining items were intended to answer the research questions, with brief explanations provided in the “comments” column. Before the formal data extraction phase, we conducted a pilot study to calibrate our data extraction process. 10 GL sources were randomly selected, and three graduate student researchers independently extracted data using the predefined data extraction form. Subsequently, we held a calibration meeting to compare and discuss all discrepancies in the extraction results. After conducting a pilot study and refining the extraction guidelines, we formally initiated the data extraction process for the remaining literature. To ensure the accuracy and objectivity of the data, we employed a strict protocol in which one researcher performed the extraction and another independently verified it. For each GL, data extraction was first conducted by one graduate student and then independently verified by a second graduate student for all extracted data items. In cases of disagreement, the two students initially attempted to reach a consensus through discussion. If the disagreement could not be resolved, the issue was submitted to the regular team meetings with the supervisor, where it was discussed by the entire team and a final decision was made, ensuring that every data item underwent thorough verification.

\subsection{Data synthesis}
In this GLR, most of the extracted data were qualitative in nature, including definitions, characteristics, quality attributes, technical elements, opportunities and challenges of AI-native applications. To analyze these data while retaining their contextual richness, we adopted Thematic Analysis, a widely used method in qualitative research for identifying, analyzing, and reporting meaningful patterns (themes) within datasets. Our process strictly followed Braun and Clarke’s six-phase framework \cite{cruzes2011recommended}: (1) familiarization with the data, (2) generating initial codes, (3) searching for themes, (4) reviewing themes, (5) defining and naming themes, and (6) reporting. During the coding process, we employed open coding \cite{corbin2014basics} for data extraction. Based on the 10 GL sources used in the data extraction phase, we first conducted a pilot data synthesis. We independently performed thematic analysis on these GL sources and then discussed the analysis results. Next, two graduate students synthesized the remaining GL sources, followed by a cross-check. Finally, the review team jointly examined the synthesis results, resolving any discrepancies through discussion and consensus. Table~\ref{tabExample} presented the thematic analysis for AI-Native application definitions, including the identified themes, candidate themes, and example quotations. All data were derived strictly from the original content of the selected GL sources, without any subjective interpretation. The replication of this GLR was made publicly available on GitHub \footnote{https://github.com/llc202jy/GLR-on-AI-native-applications}.

\begin{table*}[htbp]
\caption{Data extraction form.}
\label{tab3}
\begin{tabularx}{\textwidth}{llX}
\hline
\textbf{Data}         & \textbf{RQ} & \textbf{Comments}                                              \\ \hline
Title                 & Metadata    & N/A                                                            \\
Author                & Metadata    & N/A                                                            \\
Category              & Metadata    & Blogs/news/wiki   articles/Q\&A posts/project documents        \\
Publication   sources & Metadata    & Publication name                                               \\
Publication year      & Metadata    & N/A                                                            \\
Concept               & RQ1         & The definition of   AI-native applications                     \\
Characteristics       & RQ1         & List of characteristics   related to AI-native applications    \\
Quality   attributes  & RQ2         & List of QAs   related to AI-native applications                \\
Technology            & RQ3         & The technical elements for developing AI-native applications   \\
Opportunities              & RQ4         & List of opportunities                                               \\
Challenges            & RQ4         & List of   challenges                                           \\ 
Comments   &          & NA                                         \\ \hline
\end{tabularx}
\end{table*}

\begin{table*}[htbp]
\caption{Thematic analysis for AI-Native application definitions.}
\label{tabExample}
\begin{tabularx}{\textwidth}{p{3cm}p{5cm}X}
\hline
\textbf{Theme}                      & \textbf{Candidate Themes}            & \textbf{Example}                                                               \\ \hline
AI as a First Principle and Core Paradigm & AI as a Foundational Principle, Development Philosophy: Deep Integration, Development Philosophy: Paradigm Shift, Architectural Paradigm: AI-centric Architecture & "AI-native (or AI-first) refers to a system... fundamentally designed and built from the ground up with Artificial Intelligence (AI) as a core, foundational principle rather than as an add-on or afterthought." [G8], "An 'AI Native' application is one that... is designed with AI as a central and indispensable component, meaning that it could not exist without it." [G13], "AI native systems represent a paradigm shift, where AI is not merely an add-on but a foundational aspect." [G9] \\
Foundation Models and Generative AI & Core Technology: AI Models, Core Technology: LLMs  & "artificial intelligence as a first principle; Large Language Models (LLMs) and Generative AI" [G40], "AI-native applications represent a fundamental shift in software architecture, being built around advanced AI   models..." [G42], "foundational AI capabilities; underlying models; feedback loops; GenAI" [G31]  \\
Agentic and Orchestrated Architecture  & Architectural Paradigm: AI Orchestration, Architectural Paradigm: AI-centric Architecture, System Behavior: Autonomous Adaptation, Key Characteristics: Advanced Reasoning & "GenAI Orchestration; LLM Ops", "Designed from the ground up, AI-driven decision making, Real-time adaptation" [G2], "efficient AI models with deep SE knowledge, expert-level coding skills, and advanced reasoning capabilities" [G47]                                                             \\
Context and Memory System & Key Characteristics: Data-Driven Intelligence, System Behavior: Continuous Learning, Key Characteristics: Context-Awareness,   Quality Attributes: Trustworthiness & "AI, especially large language models; data; context;" [G34],   "trustworthy AI capabilities; data-driven and knowledge-based " [G79]  \\
Multi-modal Interaction & Key Characteristics: Multi-Modality & "These applications can interpret text, images, documents, videos, and screen data in a human-like way, enabling powerful and specialised functionalities..." [G42]                                                                                              \\ \hline
\end{tabularx}
\end{table*}

\section{Results}
\label{sec4}
\begin{figure}
\centering
\includegraphics[width=1.0\linewidth]{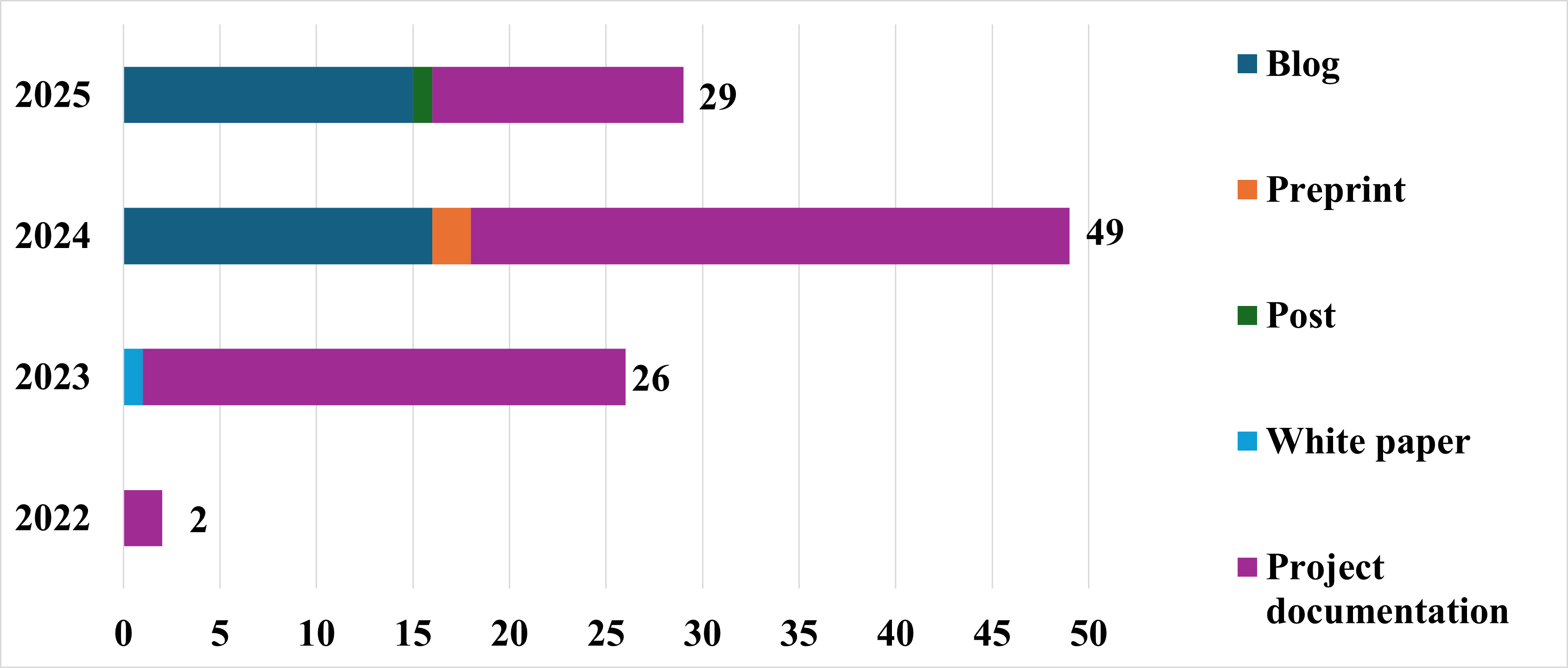}
\caption{Distribution of reviewed GL sources per year (Dec 2022 – May 2025).}
\label{demography}
\end{figure}
Based on the data extracted and synthesized from the reviewed GL sources, we addressed the research questions (cf.  Section~\ref{RQs}). This section first presents the descriptive metadata of the selected GL sources, followed by a report of the main findings related to the research questions. After completing the screening and selection process, this study identified 106 GL sources related to AI-native applications for review. Figure~\ref{demography} illustrates the annual distribution of the selected publications within the search period. Notably, in December 2022, only 2 GL sources addressed AI-native applications, which may be attributed to the fact that the concept had only just emerged during that period. Since 2023, research in this domain has increased substantially, with 26 GL sources published that year. By 2024, the number of GL sources (49) nearly doubled compared to 2023. Furthermore, in the first half of 2025 alone, 29 relevant GL sources have already been published, which accounts for nearly 60\% of the total for 2024, indicating a rapid upward trend in research on AI-native applications.

\subsection{Key characteristics (RQ1)}
\label{RQ1}
\begin{table}[]
\centering
\caption{The core elements related to the definition of AI-native applications extracted from GL sources.}
\label{Tabfig3}
\begin{tabularx}{\columnwidth}{XX} \hline
\textbf{Themes}                           & \textbf{GL sources}  \\ 
\hline
Multi-modal Interaction                   & [G42]                                        \\
Context and Memory System                 & [G31], [G34], [G79]                                                                                                    \\
Agentic and Orchestrated Architecture     & [G2], [G10], [G11], [G15], [G47]                                                                                      \\
Foundation Models and Generative AI       & [G1], [G31], [G34], [G35], [G36], [G39], [G40], [G42], [G47]                                                          \\
AI as a First Principle and Core Paradigm & [G1], [G3], [G4], [G6], [G7], [G8], [G9], [G10], [G11], [G12], [G13], [G14], [G15], [G16], [G30], [G32], [G33], [G37], [G39], [G40], [G44], [G45], [G46] 
\\ 
\hline
\end{tabularx}
\end{table}
In the review process, delineating the definition of AI-native applications was particularly crucial. Given that the term AI-native is still evolving, it may refer either to capabilities or to products. Therefore, literature that claims to define AI-native but actually describes objects such as applications, systems, or software is also considered providing a definition of AI-native applications.
\begin{figure}
\centering
\includegraphics[width=1.0\linewidth]{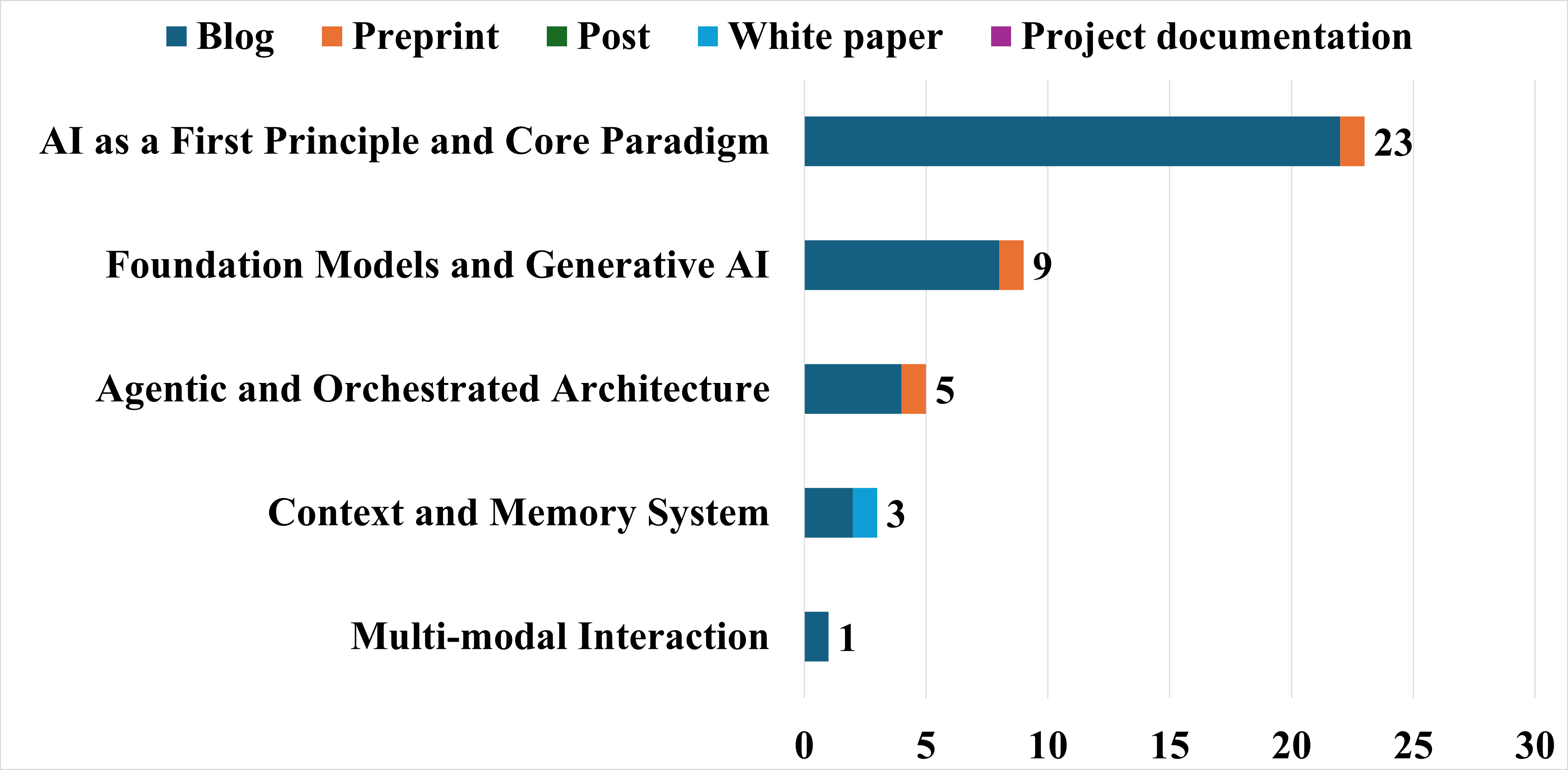}
\caption{The core elements related to the definition of AI-native applications.}
\label{fig3}
\end{figure}

We identified the core elements of the definition of AI-native applications from the reviewed GL sources, as shown in Table~\ref{Tabfig3}. Figure~\ref{fig3} illustrates the core elements included in the current definitions of AI-native applications, along with their frequencies of occurrence. The figure highlights that AI-native applications represent not only a technological innovation but also a disruption of architectural paradigms. As the most frequently occurring core element, \textbf{AI as a First Principle and Core Paradigm} clearly distinguishes between AI-native applications and AI-enabled applications that use AI as an auxiliary feature~\cite{felderer2021quality,fui2023generative}. This element emphasizes that projects must follow the principle of AI First from the very inception. The application’s architecture, functionality, workflows, and even strategic vision are all designed around AI’s capabilities and characteristics. As noted by [G42], AI-native applications represent a fundamental shift in software architecture. \textbf{Foundation Models and Generative AI} provide the technological foundation for realizing AI-native applications. In the current context, the intelligence of AI-native applications primarily stems from foundation models and generative AI technologies, exemplified by LLMs. These advanced LLMs, with their powerful capabilities in generation, comprehension, and reasoning, drive the core functionalities of AI-native applications. To fully unleash the potential of foundational models, AI-native applications adopt an orchestrated architecture [G2]. The \textbf{Agentic and Orchestrated Architecture} reveals the typical architectural paradigm of AI-native applications. In this architecture, AI assumes one or more roles, as a single agent or multiple agents, with the capability to autonomously plan, invoke tools, and collaborate with other system components, enabling the dynamic completion of complex tasks. The behavior of the application is no longer a simple input-output process but manifests as goal-driven, high-level intelligence. To deliver a coherent and personalized experience, AI-native applications must have a system capable of understanding and maintaining long-term context, along with an appropriate memory mechanism. Through feedback loops [G31], the applications can learn from past interactions and perform adaptive optimization. Without a \textbf{Context and Memory System}, AI-native applications degrade into one-off Q\&A tools. \textbf{Multi-modal Interaction} defines the human–computer interaction paradigm of AI-native applications. It is no longer limited to a single text-based interaction, but inherently supports and understands multiple data modalities, such as text, images, and audio. [G42] emphasizes that AI-native applications can comprehend and process these unstructured contents in a human-like manner, thereby providing users with a more natural and richer interaction experience. Overall, AI-native applications follow AI as a first principle and core paradigm, are driven by foundation models and generative AI, realized through agentic and orchestrated architecture, and demonstrate their value based on context and memory systems, and multi-modal interaction.

\begin{table}[]
\centering
\caption{The key characteristics of AI-native applications extracted from GL sources.}
\label{Tabfig4}
\begin{tabularx}{\columnwidth}{p{3cm}X} \hline
\textbf{Themes}                           & \textbf{GL sources}  \\ 
\hline
Continuous Learning and Adaptability           & [G7], [G10], [G14], [G16], [G30], [G31], [G39], [G79]                                             \\
Enhanced Developer Experience and Tooling     & [G4], [G47], [G48], [G50], [G52], [G55], [G57], [G60], [G64], [G84], [G87], [G104]                                                                 \\
Ecosystem Integration and Flexibility          & [G19], [G48], [G51], [G54], [G71], [G72], [G78], [G80], [G83], [G92], [G93], [G94], [G105], [G106]                                                   \\
Trust, Privacy, and Security                   & [G17], [G20], [G21], [G22], [G26], [G27], [G49], [G51], [G56], [G61], [G62], [G63], [G72], [G86], [G88], [G89], [G91], [G93], [G99]                                                                   \\
Intelligent and Adaptive Interfaces           & [G1], [G13], [G19], [G21], [G23], [G24], [G28], [G34], [G35], [G36], [G41], [G44], [G58], [G63], [G78], [G80], [G81], [G83], [G88], [G91], [G101]                                                    \\
Autonomous Systems and Agentic Behavior       & [G8], [G11], [G13], [G15], [G20], [G23], [G26], [G36], [G37], [G40], [G42], [G46], [G50], [G52], [G53], [G65], [G67], [G73], [G75], [G81], [G89], [G94], [G95], [G97], [G102]                                                                           \\
Generative and Multimodal Capabilities        & [G2], [G18], [G22], [G25], [G27], [G28], [G29], [G54], [G56], [G60], [G64], [G66], [G69], [G71], [G74], [G76], [G77], [G82], [G85], [G86], [G90], [G96], [G98], [G99], [G100], [G106]                                                                \\
AI-Centric Architecture and Core Capabilities & [G1], [G2], [G6], [G7], [G8], [G9], [G10], [G11], [G14], [G15], [G16], [G18], [G24], [G30], [G31], [G32], [G33], [G37], [G39], [G40], [G42], [G43], [G44], [G46], [G55], [G58], [G59], [G62], [G66], [G68], [G70], [G73], [G79], [G87], [G102], [G103], [G104], [G105] 
\\ 
\hline
\end{tabularx}
\end{table}

\begin{figure}
\centering
\includegraphics[width=1.0\linewidth]{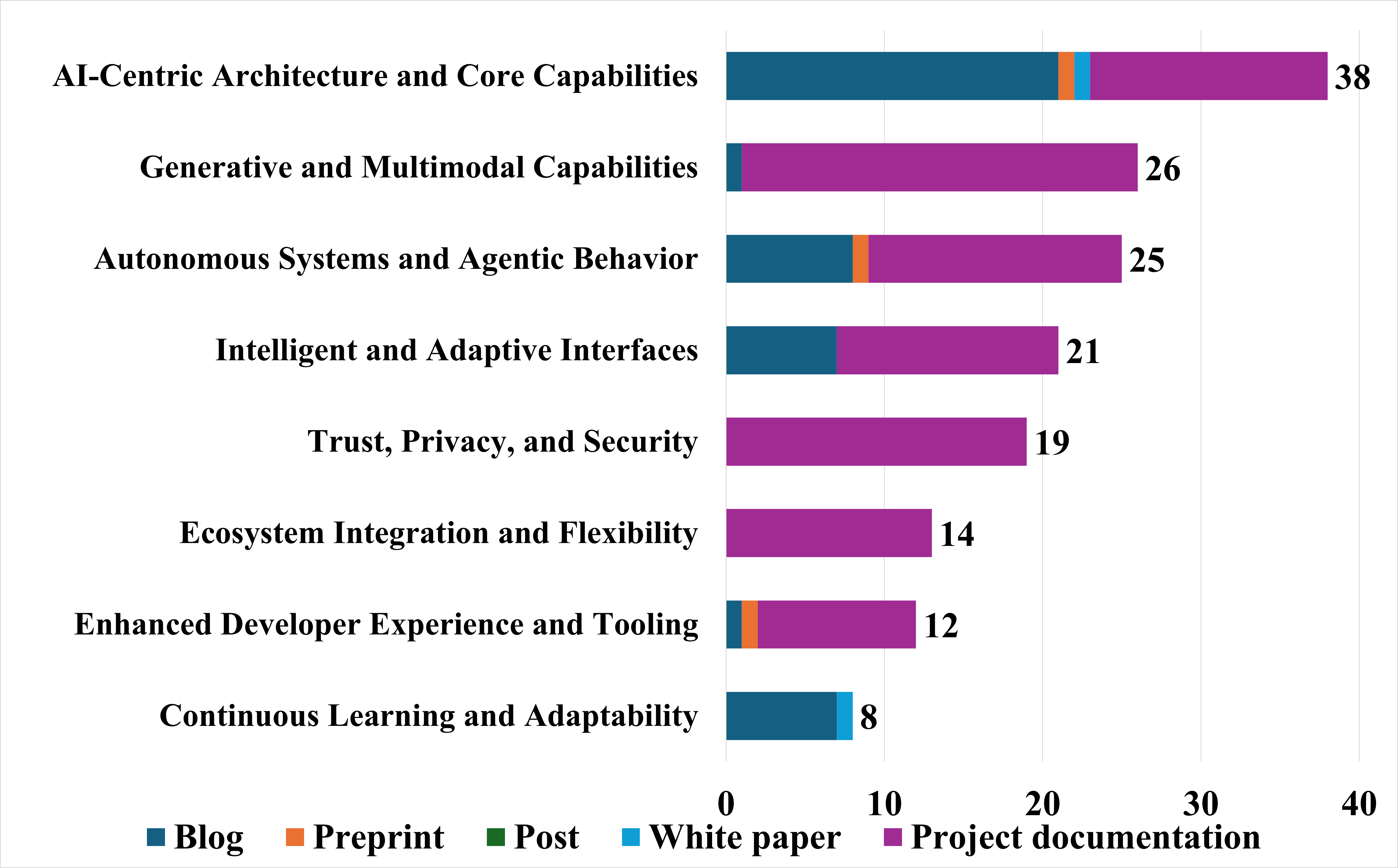}
\caption{The key characteristics of AI-Native applications.}
\label{fig4}
\end{figure}
Building on the clarified core elements of the definition of AI-native applications, we further identified and analyzed the key characteristics of AI-native applications, as shown in Table~\ref{Tabfig4}. \textbf{AI-Centric Architecture and Core Capabilities} describe the fundamental construction of AI-native applications, where AI models serve as the core logical units [G15], following a data-first and model-first design philosophy. It emphasizes that AI is not bolted on as an afterthought but is intertwined with the application’s design [G1], ensuring that intelligence is an intrinsic property. As shown in Figure~\ref{fig4}, this characteristic represents the most fundamental feature of AI-native applications. \textbf{Intelligent and Adaptive Interfaces} describe the transformation of human–computer interaction and emphasize the dynamically evolving nature of AI-native applications. They are capable of continuously improving through ongoing interactions. For example, the concept of generative UI is discussed in [G13, G83], implying that user interfaces are no longer predefined but can be generated in real time by AI based on user intent and context. This enables dynamic conversational interactions and generative interfaces [G19], providing users with seamless and highly personalized experiences. AI-native applications are designed to continuously learn and improve [G30], which is achieved through built-in mechanisms, such as feedback loops that leverage real-world data to optimize performance [G31]. They are able to learn from each interaction, thereby becoming more adaptive and responsive to changing environments [G39]. These characteristics enable AI-native applications to exhibit dynamic behaviors based on real-time inputs and memory, continuously evolving. \textbf{Continuous Learning and Adaptability} encompass the proactiveness and task execution capabilities of AI-native applications, allowing them to autonomously accomplish complex tasks as intelligent agents. The goal of \textbf{Autonomous Systems and Agentic Behavior} is to minimize human intervention by achieving zero-touch automation through intelligent agents [G37]. These agents can autonomously take actions, perform multi-step task planning, and invoke tools, marking a shift of AI-native applications from passive tools to proactive actors. \textbf{Generative and Multimodal Capabilities} indicate that the capabilities of AI-native applications go beyond text generation. They possess strong multimodal abilities, enabling content understanding and generation across multilingual text, images, and videos [76, 96]. Moreover, they can interact with various input modalities such as speech recognition [56] and text-to-speech [74]. To cope with the complexity of AI-native applications, a new developer ecosystem is emerging. To lower the barriers to entry and improve development efficiency, specialized toolchains and frameworks have been introduced, such as LLM orchestration frameworks including LangChain, Semantic Kernel, and Haystack [G15]. \textbf{Enhanced Developer Experience and Tooling} Enhanced Developer Experience and Tooling aim to promote a programming paradigm that facilitates human–AI collaborative development, making the development process itself more intelligent and efficient. \textbf{Ecosystem Integration and Flexibility} indicate that AI-native applications are designed to be vendor-agnostic [G83] and support broad integration. They can connect to external tools via the Model Context Protocol [G19], thereby dynamically extending their functional boundaries. Notably, \textbf{Trust, Privacy, and Security} emphasize the critical quality attributes of AI-native applications. For AI-native applications, privacy is the highest-priority consideration, with the core principle being privacy by design [G20]. 

The concept and boundaries of AI-native applications are jointly defined by the eight core characteristics described above. Built upon an AI-centric architecture, they interact through intelligent and adaptive interfaces, are driven by autonomous agent systems with continuous learning capabilities, and enable generative and multimodal creation. The realization of the AI-native application paradigm relies on enhanced developer tooling and flexible ecosystem integration, and must fundamentally be safeguarded by trust, privacy, and security to ensure its healthy evolution.

\begin{tcolorbox}[colback=gray!20, colframe=gray!20]
\textbf{Summary to answer RQ1.} This study identifies 5 core elements related to the definition of AI-native applications and 8 key characteristics of AI-native applications are highlighted.  The primary, non-negotiable characteristic of an AI-native application is that AI serves as the core pillar of its architecture and value proposition, rather than being an embedded or supplementary feature.
\end{tcolorbox}

\subsection{Quality attributes (RQ2)}
\label{RQ2}
\begin{figure}
\centering
\includegraphics[width=1.0\linewidth]{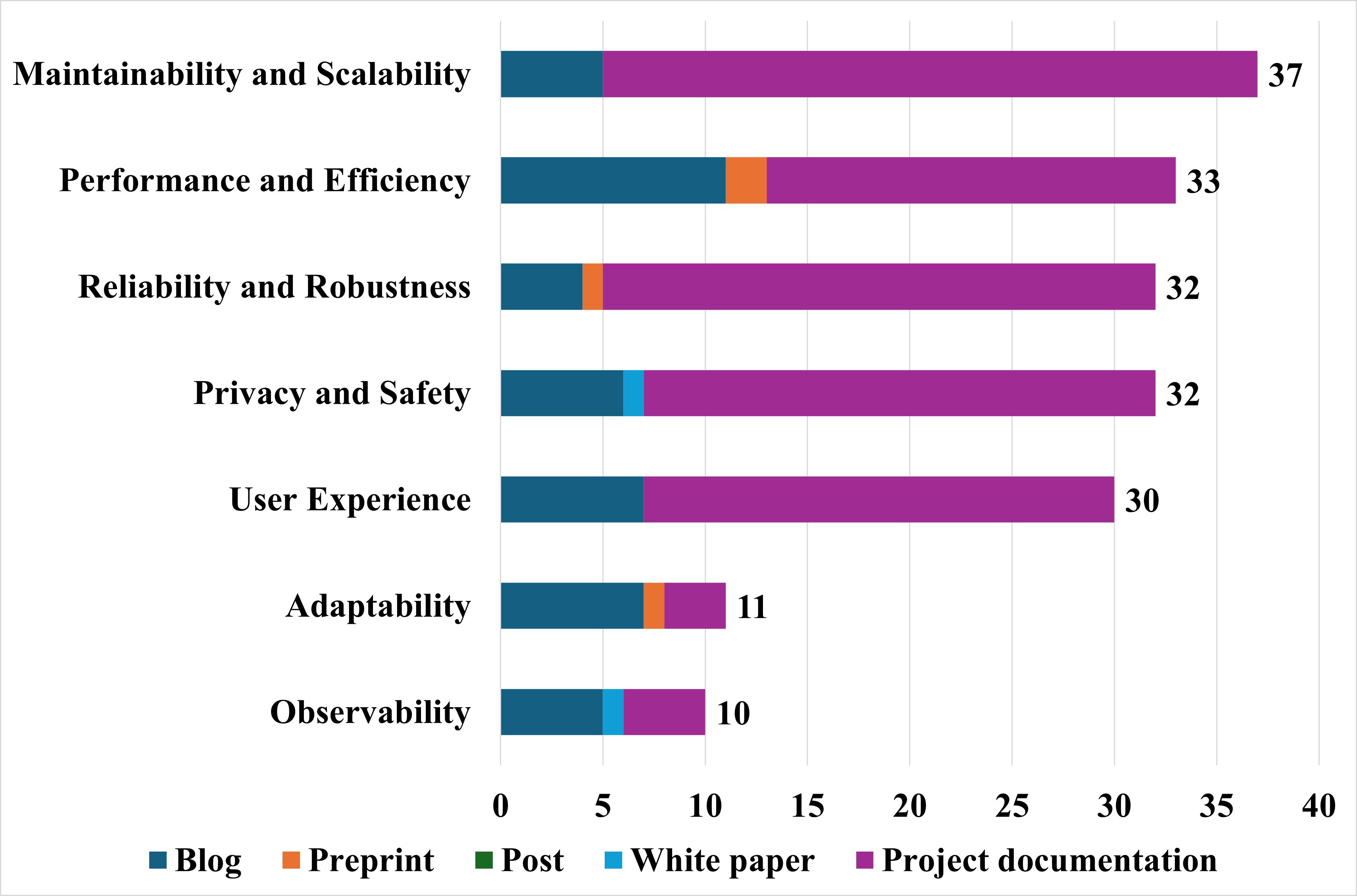}
\caption{The concerned quality attributes for AI-native applications.}
\label{fig5}
\end{figure}

The analysis of the grey literature reveals a distinct hierarchy of quality attributes critical to the success of AI-native applications. While many of these attributes are well established in traditional software engineering, their prioritization, interpretation, and interdependencies are fundamentally reshaped by the central role of AI. We identified concerned quality attributes for AI-native applications from the reviewed GL sources, as shown in Table~\ref{Tabfig5}. At the forefront of developer concerns are attributes that form the foundation of any robust software system. \textbf{Maintainability and Scalability} emerged as the most frequently cited quality attribute, closely followed by \textbf{Performance and Efficiency} and \textbf{Reliability and Robustness} (see Figure~\ref{fig5}). This indicates that the AI-native application paradigm does not abandon traditional software engineering principles but rather amplifies their significance.

\begin{table}[]
\centering
\caption{The concerned quality attributes for AI-native applications extracted from GL sources.}
\label{Tabfig5}
\begin{tabularx}{\columnwidth}{p{2cm}X} \hline
\textbf{Themes}                           & \textbf{GL sources}  \\ 
\hline
Observability                   & [G6], [G10], [G15], [G31], [G38], [G55], [G59], [G65], [G79], [G91]                                                   \\
Adaptability                    & [G8], [G9], [G14], [G16], [G24], [G32], [G39], [G42], [G47], [G73], [G98]                                                   \\
User Experience                  & [G2], [G6], [G8], [G14], [G17], [G18], [G19], [G23], [G26], [G30], [G34], [G35], [G50], [G51], [G55], [G56], [G58], [G62], [G63], [G64], [G65], [G75], [G76], [G80], [G81], [G83], [G86], [G87], [G88], [G95]                                                                                                                                    \\
Privacy and Safety               & [G1], [G4], [G10], [G11], [G19], [G20], [G21], [G22], [G26], [G27], [G39], [G40], [G49], [G51], [G53], [G58], [G59], [G60], [G62], [G63], [G71], [G72], [G79], [G80], [G82], [G83], [G87], [G89], [G93], [G99], [G100], [G105]                                                                                  \\
Reliability and Robustness       & [G2], [G4], [G15], [G24], [G25], [G27], [G41], [G46], [G50], [G56], [G57], [G61], [G64], [G66], [G68], [G69], [G70], [G76], [G77], [G78], [G82], [G84], [G85], [G90], [G92], [G94], [G97], [G98], [G100], [G101], [G102], [G103]                                                                                  \\
Performance and Efficiency       & [G1], [G3], [G7], [G11], [G18], [G20], [G28], [G29], [G30], [G31], [G32], [G37], [G38], [G40], [G43], [G46], [G47], [G48], [G49], [G52], [G57], [G60], [G67], [G71], [G73], [G74], [G84], [G85], [G86], [G92], [G102], [G104], [G106]                                                                                 \\
Maintainability and Scalability  & [G3], [G7], [G9], [G17], [G21], [G22], [G23], [G25], [G28], [G29], [G37], [G43], [G48], [G52], [G53], [G61], [G67], [G69], [G72], [G74], [G75], [G77], [G78], [G81], [G88], [G89], [G90], [G91], [G93], [G95], [G97], [G99], [G101], [G103], [G104], [G105], [G106]
\\ 
\hline
\end{tabularx}
\end{table}

\textbf{Maintainability and Scalability} relate to the architectural quality of the system, referring to the ability to transform AI models into mature software products. They focus on the system's integration complexity, maintenance cost, and capability to handle business growth. This attribute is oriented towards DevOps/LLMOps, ensuring that the technology stack can support business expansion and integrate with the existing IT ecosystem. \textbf{Performance and Efficiency} describes the operational effectiveness of AI systems under resource constraints. It focuses on the balance between response speed, computational cost, and output value. Due to the high inference cost of LLMs,"Cost- and energy-effectiveness" [G43] and "Resource consumption" [G38] have become decisive factors in determining the commercial feasibility (ROI) of applications [G37]. \textbf{Reliability and Robustness} refers to the ability of AI systems to continuously deliver accurate, truthful, and consistent outputs across diverse environments and input conditions. This is a core quality attribute for addressing the issue of AI hallucinations and ensuring the correctness of business logic, and it also represents a major pain point currently faced by AI-native applications. The frequently mentioned Accuracy [G2, G46, G57] and Hallucination [G15, G90] indicate that users place extremely high demands on the fidelity of model outputs. \textbf{Privacy and Safety} refers to the capability of ensuring that AI systems operate within ethical, legal, and safety boundaries. It focuses on preventing systems from causing harm to users, misusing data, or generating harmful content, and serves as a prerequisite for users’ adoption of AI-native applications. Unlike traditional security, this attribute places greater emphasis on the behavioral safety of AI models, such as biased outputs [G1].   \textbf{User Experience} focuses on the quality of human-AI interaction. In AI-native applications, the core of UX has shifted from click flows to conversational flows. Users expect the system not only to provide Usability [G56, G58], but also Personalization [G2, G14]. Natural language queries lower the barrier to use, but also require the system to possess extremely high intent recognition capabilities. \textbf{Adaptability} reflects the unique intelligent characteristics of AI-native applications, namely the ability of the system to self-optimize, continuously learn, and accomplish complex tasks with minimal human intervention. This is a key attribute distinguishing traditional software from AI-native applications. While traditional software is static, AI applications are expected to possess continuous learning and self-evolving capabilities [G8, G32]. In AI-native applications, \textbf{Observability} refers to the ability to gain insight into the internal states, decision logic, and operational metrics of AI black boxes, serving as the foundation for system debugging, trust building, and continuous optimization. Due to the inherent opacity of large models, observability becomes crucial. Analyses indicate that developers need not only to monitor traditional system metrics but also to employ "Logging prompt+context+output" [G15] to review model behavior. Transparency [G10] and Explainability [G31] serve as the bridge connecting technical implementation with user trust.

These attributes represent the fundamental expectations, whose importance is heightened by the new complexities introduced by AI.  While foundational attributes remain essential, the AI-native context reshapes their internal priorities and introduces new dimensions. These attributes are interrelated and mutually constraining, jointly determining the overall quality of AI-native applications. A lack of transparency and explainability makes it difficult for users to establish trust in the AI system; in the pursuit of extremely low response latency and operational costs, trade-offs often need to be made in model scale or inference depth, which may affect the accuracy of results. Meanwhile, achieving a highly personalized user experience requires the system to have continuous perception, adaptation, and learning capabilities regarding user behavior. Without good maintainability, these design goals in performance, security, and user experience will be difficult to realize in a stable and long-term manner in production environments. The results indicate a shift from a purely deterministic conception of software quality toward one that embraces the probabilistic and data-driven nature inherent to AI-native applications.

\begin{tcolorbox}[colback=gray!20, colframe=gray!20]
\textbf{Summary to answer RQ2.} This study identified 7 quality attributes most prominently considered during the design of AI-native applications: Maintainability and Scalability, Performance and Efficiency, Reliability and Robustness, Privacy and Safety, Usability, Adaptability and Observability.  At the current stage, mitigating hallucinations and reducing costs are the most pressing concerns for developers; however, in the long run, a system’s adaptability and observability will determine the lifecycle and performance ceiling of AI-native applications.
\end{tcolorbox}

\subsection{Technical elements (RQ3)}
We identified the development technology elements of AI-native applications from the reviewed GL sources, as shown in Table~\ref{Tabfig7}. These technical elements outline the current technological landscape of AI-native application development. They center on core model technologies and are customized through model training. Developers enhance their capabilities using RAG and agentic systems, deploy them via infrastructure, and provide access through local AI or cloud interfaces. Finally, by combining them with generative media capabilities, developers assemble these components into user-facing products through user interface and application frameworks.

\begin{figure}
\centering
\includegraphics[width=1.0\linewidth]{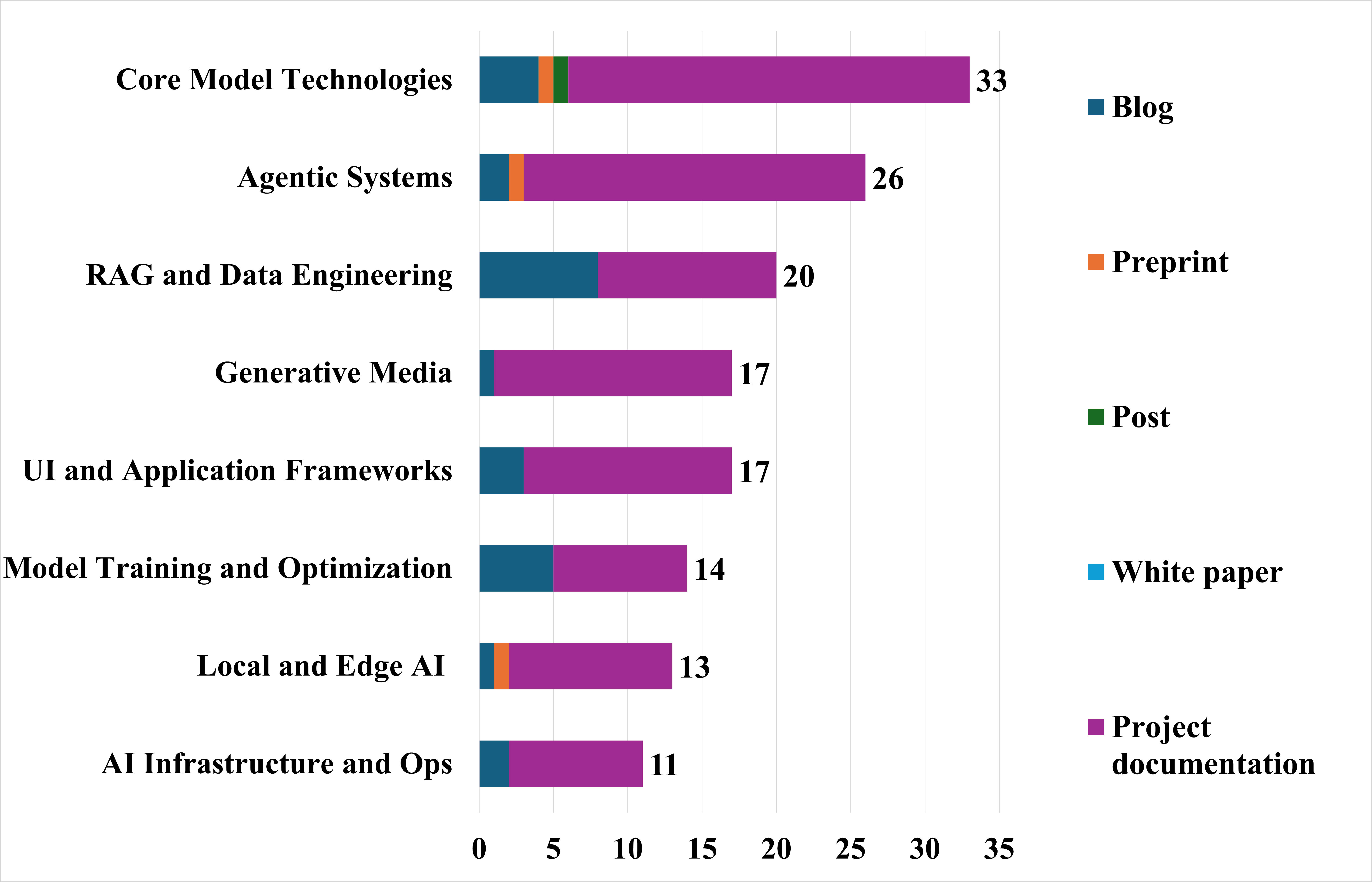}
\caption{The technical elements used in the development of AI-native applications.}
\label{fig7}
\end{figure}

\begin{table}[]
\centering
\caption{The technical elements used in the development of AI-native applications extracted from GL sources.}
\label{Tabfig7}
\begin{tabularx}{\columnwidth}{p{2.5cm}X} \hline
\textbf{Themes}                           & \textbf{GL sources}  \\ 
\hline
AI Infrastructure and Ops        & [G18], [G19], [G21], [G35], [G38], [G50], [G58], [G64], [G80], [G82], [G84]  \\
Local and Edge AI                & [G1], [G17], [G23], [G26], [G47], [G49], [G51], [G58], [G62], [G64], [G81], [G99], [G104]\\
Model Training and Optimization  & [G6], [G9], [G28], [G31], [G37], [G41], [G61], [G63], [G71], [G85], [G90], [G96], [G99], [G101]  \\
UI and Application Frameworks            & [G15], [G32], [G35], [G48], [G49], [G50], [G51], [G54], [G59], [G60], [G70], [G81], [G83], [G84], [G86], [G87], [G88]                                                     \\
Generative Media                 & [G22], [G25], [G29], [G30], [G54], [G56], [G63], [G66], [G69], [G70], [G72], [G75], [G77], [G78], [G86], [G90], [G100]  \\
RAG and Data Engineering         & [G2], [G10], [G12], [G15], [G17], [G18], [G32], [G34], [G41], [G43], [G55], [G57], [G62], [G71], [G73], [G82], [G88], [G89], [G98], [G102] \\
Agentic Systems                  & [G6], [G10], [G19], [G20], [G21], [G22], [G23], [G24], [G26], [G47], [G48], [G52], [G53], [G55], [G56], [G57], [G65], [G67], [G91], [G92], [G93], [G94], [G95], [G102], [G104], [G105]  \\
Core Model Technologies          & [G1], [G5], [G12], [G24], [G27], [G30], [G37], [G46], [G52], [G59], [G65], [G66], [G67], [G68], [G69], [G72], [G73], [G74], [G75], [G76], [G78], [G80], [G87], [G89], [G91], [G92], [G94], [G96], [G97], [G98], [G101], [G103], [G106]
\\ 
\hline
\end{tabularx}
\end{table}
As shown in Figure~\ref{fig7}, \textbf{Core Model Technologies} form the foundational brain of AI-native applications, encompassing the core algorithms, model architectures, and base large models that drive application intelligence. They focus not on how models are deployed or applied, but on the types of models and the underlying technical principles. Developers no longer rely solely on a single closed-source large model. Instead, models have become diversified and specialized; in addition to traditional LLMs, Transformer architectures are now widely applied to visual and sequential data processing [G24, G66]. \textbf{Agentic Systems} represent the evolution of AI applications from chat-based interfaces to “autonomous agents.” They focus on how AI plans tasks, uses tools, executes complex workflows, and interacts with other systems. Notably, MCP has emerged as an important new standard [G91, G105], connecting AI models with data sources or tools to address agent interoperability. \textbf{RAG and Data Engineering} involve the engineered processes by which AI models connect to external knowledge bases, as well as handle, store, and retrieve data to enhance model capabilities. Currently, vector databases [G15] and graph databases [G2] serve as the physical carriers of memory. This technical element is a key stack for mitigating model hallucinations and enriching domain knowledge. \textbf{Generative Media} focuses on modalities beyond text, including technologies for generating and recognizing images, videos, and speech, such as Stable Diffusion [G86], Whisper [G90], and FLUX [G69]. It forms the core of multimodal interactive experiences. \textbf{UI and Application Frameworks} cover the traditional software engineering elements needed to build AI-native applications, namely the user interfaces, business logic, and programming frameworks that wrap around the AI model. Next.js and React [G70, G81] are mainstream choices for developing the front end of AI applications, offering rapid UI iteration capabilities. \textbf{Model Training and Optimization} involves the process of creating or modifying models, focusing on improving specific performance or efficiency through data and algorithms. Fine-tuning [G28, G31] is the core approach for adapting models to specific business scenarios. \textbf{Local and Edge AI} focuses on running AI models directly on user-end devices, aiming to address issues related to privacy, latency, and offline usage. The popularity of tools like Ollama and llama.cpp [G51, G62, G99] enables developers to easily deploy models on local laptops or mobile devices. \textbf{AI Infrastructure and Ops} concerns the full lifecycle runtime environment of AI applications, from scheduling cloud computing resources to monitoring and maintaining applications after deployment. Docker and Kubernetes are the foundational technologies for deploying AI services [G21, G84].

The development of AI-native applications has evolved beyond simply invoking large model APIs into a sophisticated systems engineering endeavor. It demands that developers combine traditional full-stack engineering skills with emerging AI-native capabilities, such as RAG and Agents. Future AI applications are expected to be highly autonomous, deeply context-aware, and seamlessly integrated into every device.

\begin{tcolorbox}[colback=gray!20, colframe=gray!20]
\textbf{Summary to answer RQ3.} This study identified 8 technical elements used in the development of AI-native applications: Core Model Technologies, Agentic Systems, RAG and Data Engineering, Generative Media, UI and Application Frameworks, Model Training and Optimization, Local and Edge AI, and AI Infrastructure and Ops. These elements encompass foundational model technologies, as well as higher-level agent systems, data engineering, and the engineering infrastructure that supports real world deployment and application.
\end{tcolorbox}

\subsection{Opportunities and challenges (RQ4)}
We identified the opportunities and challenges of AI-native applications from the reviewed GL sources, as shown in Table~\ref{Tabfig9} and ~\ref{Tabfig10}. 
\begin{table}[]
\centering
\caption{The opportunities presented by AI-native applications extracted from GL sources.}
\label{Tabfig9}
\begin{tabularx}{\columnwidth}{p{2cm}X} \hline
\textbf{Themes}                           & \textbf{GL sources}  \\ 
\hline
Democratization        & [G30], [G52], [G54], [G56], [G70], [G105]    \\
Edge and Privacy       & [G1], [G23], [G27], [G49], [G51], [G62], [G81], [G89], [G93]  \\
Strategic Value        & [G7], [G8], [G9], [G15], [G33], [G37], [G38], [G39], [G40], [G61], [G82], [G83], [G87], [G98]  \\
Personalization        & [G2], [G8], [G10], [G14], [G15], [G16], [G17], [G20], [G30], [G32], [G40], [G50], [G62], [G71], [G101]  \\
AI-Driven Development  & [G11], [G19], [G44], [G46], [G47], [G53], [G55], [G57], [G59], [G64], [G72], [G74], [G84], [G87], [G94], [G95], [G97], [G104], [G105]     \\
Data Intelligence      & [G6], [G14], [G16], [G18], [G20], [G21], [G31], [G34], [G42], [G45], [G48], [G66], [G68], [G73], [G80], [G82], [G88], [G89], [G93], [G102], [G103]        \\
Autonomous Agents      & [G2], [G3], [G6], [G9], [G11], [G13], [G19], [G21], [G22], [G24], [G26], [G36], [G39], [G44], [G50], [G53], [G58], [G65], [G67], [G75], [G91], [G99]      \\
Multimodal Interaction & [G12], [G13], [G17], [G22], [G25], [G27], [G28], [G29], [G31], [G34], [G36], [G49], [G54], [G63], [G64], [G65], [G69], [G72], [G76], [G77], [G78], [G85], [G86], [G90], [G92], [G96], [G98], [G100], [G101], [G106]        \\ 
\hline
\end{tabularx}
\end{table}

\begin{figure}
\centering
\includegraphics[width=1.0\linewidth]{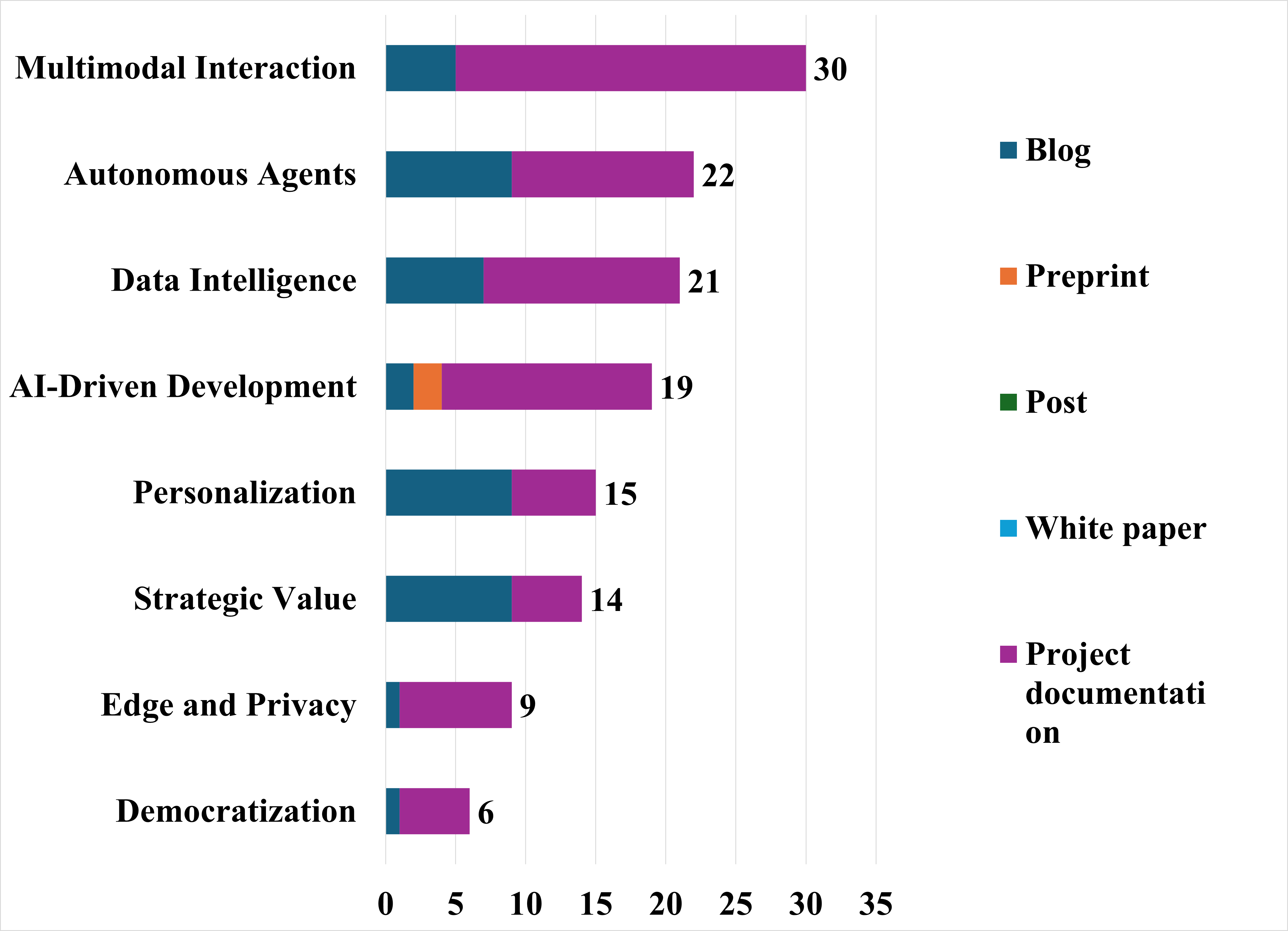}
\caption{The opportunities presented by AI-Native applications.}
\label{fig9}
\end{figure}

The opportunities presented by AI-native applications are multi-faceted (see Figure~\ref{fig9}). The foremost opportunity, is \textbf{Multimodal Interaction}. This opportunity marks the dissolution of the boundaries between humans and machines. Systems are no longer limited to single text-based commands; instead, they communicate with users naturally and seamlessly through voice [G85], vision, video [G77], and even dynamically generated interfaces [G64]. This integration of digital experiences into the physical world greatly lowers cognitive barriers and reduces friction in interaction. The significant emphasis on the \textbf{Autonomous Agents}  reveals a critical insight, the agency of software is awakening. AI applications are evolving from passive tools that wait for instructions into intelligent entities capable of autonomously planning paths, invoking tools, and executing complex cross-application workflows based on vague goals [G36, G53, G65]. This signifies a fundamental shift in the human role in digital interactions, from being the direct operator of tedious tasks to becoming a results-oriented overseer, thereby unlocking immense productivity potential. At the same time, the core capabilities of applications are evolving from simple information retrieval to \textbf{Data Intelligence}. Analysis shows that through RAG and multimodal reasoning capabilities, AI-native applications can effectively process unstructured data such as PDFs and videos, transforming them into actionable structured knowledge [G66, G102]. These systems demonstrate the ability to extract logic from massive and chaotic information [G42], perform reasoning, and generate insights, marking a qualitative shift in data processing from storage and retrieval to understanding and reasoning. 

\textbf{AI-Driven Development} reveals a fundamental shift in the software engineering paradigm. AI is no longer merely an auxiliary tool for code completion. It has become a core builder spanning the entire lifecycle, from requirements analysis and architecture design to code generation and automated operations [G104]. This transformation significantly shortens development cycles, making it much more feasible for solo entrepreneurs or very small teams to build complex systems [G44]. \textbf{Democratization} has become the most prominent social feature of AI-native applications. Data show that through natural language interaction and low-code or no-code platforms, the power to construct complex automated workflows is being transferred from professional engineers to non-technical users [G52, G70]. This de-skilling trend not only broadens the base of innovators but also enables long-tail market demands to be met at low cost [G30]. At the same time, the infrastructure layer is showing a strong trend back toward \textbf{Edge and Privacy}. Unlike early models that relied heavily on cloud computing, AI-native applications emphasize the importance of on-device inference [G1, G27, G51]. This is not only a technical response to low-latency interactions but also a fundamental return to the needs of data sovereignty and privacy protection. Edge intelligence allows AI applications to operate in offline and sensitive data environments, alleviating users’ trust concerns about transmitting data to the cloud and laying a foundation of trust for the adoption of AI in high-privacy domains such as healthcare and personal assistants. In terms of software production and data utilization, AI’s influence has penetrated core processes.  These opportunities converge on the upgrade of user experience and the realization of business value. \textbf{Personalization} indicates that AI-native applications are moving beyond rule-based traditional recommendation systems toward a dynamic adaptation model based on long-term memory and context awareness. Systems can capture users’ behavioral patterns, preference details, and even emotional states, providing a personalized service experience with a digital twin nature [G10]. This experience exhibits increasing marginal utility as interaction time grows [G32]. From a macro perspective, all these capabilities ultimately point to \textbf{Strategic Value}. Adopting an AI-native architecture is no longer merely a tactical efficiency tool. AI-native applications, by restructuring cost structures and value creation chains, are becoming a core engine driving exponential growth in organizational ROI [G39, G40].

\begin{table}[]
\centering
\caption{The challenges presented by AI-native applications extracted from GL sources.}
\label{Tabfig10}
\begin{tabularx}{\columnwidth}{p{3cm}X} \hline
\textbf{Themes}                           & \textbf{GL sources}  \\ 
\hline
Market Maturity and Standards   & [G39], [G45]  \\
UX and Accessibility            & [G34], [G35], [G51] \\
Cost and Operations             & [G3], [G14], [G31], [G45], [G60] \\
Reliability and Trustworthiness & [G5], [G10], [G46], [G47], [G57], [G87] \\
Integration and Ecosystem       & [G17], [G34], [G38], [G42], [G57], [G75], [G84]  \\
Data Privacy and Governance     & [G1], [G3], [G10], [G14], [G31], [G46], [G71], [G87], [G88]    \\
Infrastructure and Performance  & [G1], [G17], [G35], [G38], [G42], [G47], [G51], [G71], [G75], [G88], [G103]
\\ 
\hline
\end{tabularx}
\end{table}

\begin{figure}
\centering
\includegraphics[width=1.0\linewidth]{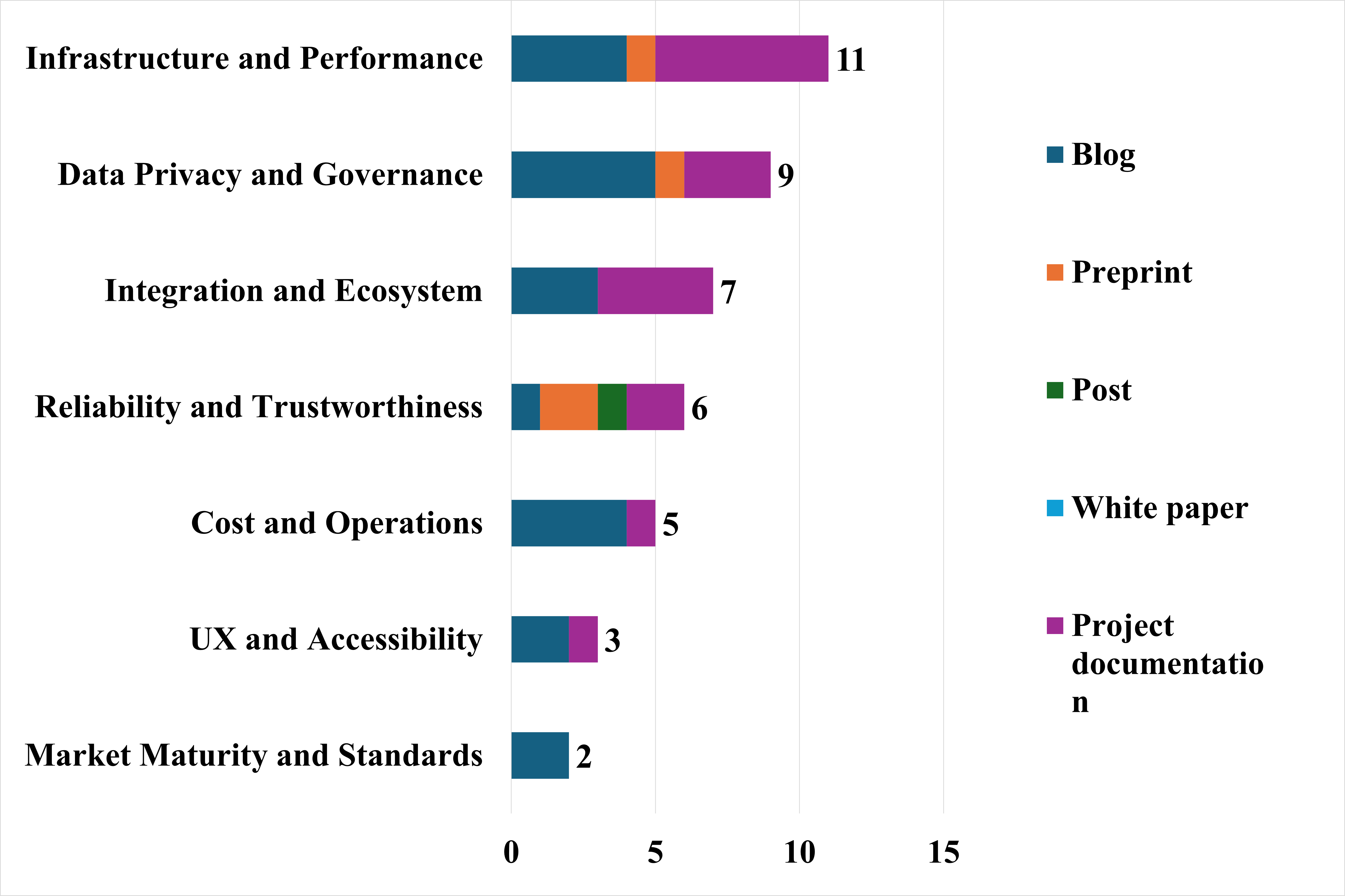}
\caption{The challenges presented by AI-Native applications.}
\label{fig10}
\end{figure}

However, these opportunities are accompanied by profound challenges. As shown in Figure~\ref{fig10}, the primary challenge is  \textbf{Infrastructure and Performance}. Although cloud computing resources are abundant, on-device deployment is constrained by physical hardware limitations, especially on mobile devices, where a zero-sum trade-off exists between computational resources and user experience. [G1] notes that hardware constraints directly create bottlenecks for complex models. Moreover, these technical bottlenecks are amplified by the inertia of existing enterprise IT architectures. AI-native applications emphasize real-time, asynchronous, and streaming interactions, which are at odds with traditional synchronous architectures. Data silos force developers to deal with fragmented cross-platform data, further exacerbating integration complexity [G34], and introducing \textbf{Integration and Ecosystem} challenge. As AI applications penetrate core business operations, \textbf{Data Privacy and Governance}, and \textbf{Reliability and Trustworthiness} have become non-negotiable red lines. Analysis shows that while on-device processing is regarded as a solution for privacy protection, it introduces new security challenges. Due to the uncontrollability of the plugin ecosystem, automatic detection mechanisms for sensitive data often fail, leading to a surge in compliance risks [G88]. At the same time, the black-box nature of AI gives rise to a crisis of explainability. Users care not only about the accuracy of results but also about the transparency of the decision-making process. When a system cannot explain its logic, legal risks and trust deficits inevitably follow [G10]. Moreover, enterprises must invest significant effort in correcting model hallucinations and biases to ensure the reliability of outputs. At the business level, \textbf{Cost and Operations} as well as \textbf{Market Maturity and Standards} constitute the practical barriers to implementation. The high costs of AI-native applications are reflected not only in initial deployment but also in ongoing token consumption and model maintenance [G60]. \textbf{UX and Accessibility} determine the breadth of technology adoption. Although AI technology is becoming increasingly complex, end users expect low-barrier interactions. Many current applications still require users to have specific knowledge of Prompt Engineering, creating significant usage barriers [G34].

\begin{tcolorbox}[colback=gray!20, colframe=gray!20]
\textbf{Summary to answer RQ4.} This study identified 8 opportunities and 7 challenges presented by AI-native applications. For developers and enterprises, the opportunity lies not only in using AI to improve the efficiency of existing processes, but also in leveraging agents and multimodal capabilities to create unprecedented product experiences. Future solutions will need to shift from optimizing individual models to systems engineering, with a focus on efficient on-device inference, data privacy and compliance, and low-cost operations and maintenance.
\end{tcolorbox}

\section{Discussion}
\label{sec5}
\subsection{Interpretation of Results}
\subsubsection{Interpretation of RQ1 Results}
AI-native applications possess eight key characteristics (cf. Section~\ref{RQ1}), which clearly distinguish AI-native applications from those that are merely AI-enabled. They exhibit dynamic properties, enabling continuous learning and evolution through interaction with data and users, thereby enhancing efficiency and personalization. In AI-native applications, AI serves as the core engine of the workflow, and removing it not only diminishes functionality but also renders the application fundamentally inert. In contrast, AI-enabled applications~\cite{martinez2022software,felderer2021quality,indykov2025architectural,gozalo2023survey} augment a deterministic workflow with AI features, yet the core functionality remains intact even without the AI component. For example, Google Search, as a mature and highly successful commercial model, is integrating generative AI to enhance its search capabilities. By incorporating AI-generated answers alongside traditional search results, Google improves user experience while maintaining the integrity of its core revenue model. By comparison, Perplexity AI  exemplifies the AI-native application paradigm, aiming to construct an answer engine rather than a traditional search engine. This approach prioritizes delivering direct answers and emphasizing user experience, rather than relying on conventional search result listings.

In the results for RQ1, a notable pattern in data source distribution emerged. High-level, conceptual definitions of AI-native applications primarily originate from technical blogs and industry articles. In contrast, detailed descriptions of the key characteristics of AI-native applications are predominantly derived from open-source project documentation on GitHub (71 out of 106 items, accounting for over 60\%). This distinction in sources is not accidental. It reveals the dynamic process through which AI-native applications evolve from theoretical conception to practical implementation. In this study, technical blogs play the role of theoretical leaders, proposing high-level concepts such as “AI-centric architecture” [G1] and “continuous learning and adaptability” [G30]. Conversely, the vast amount of data on GitHub shapes practice, concretizing these grand visions in a bottom-up, practice-driven manner. GitHub practices are not merely implementations of theory but also a materialization and selective refinement of it. For example, the theme of \textbf{Trust, Privacy, and Safety} is often mentioned broadly in blogs, but in GitHub data it is translated into a series of actionable, concrete features, such as “Privacy-First: 100\% offline processing ensures your data never leaves your device” [G63]. This indicates that the developer community regards on-device processing as the most essential and feasible approach for ensuring privacy and trust in AI-native applications. Since the dataset primarily comes from GitHub, the analysis inevitably reflects a developer-centered perspective, which is particularly evident in the theme of \textbf{Enhanced Developer Experience and Tooling}. 

It is important to note that AI-native applications do not necessarily require generative AI functionality from the outset. Just as some iconic software suites from the era of on-premises deployment successfully transitioned to cloud-native versions, such as Adobe Photoshop and Microsoft Office, we posit that many companies can, over time, gradually evolve from cloud-native to AI-native applications.

\subsubsection{Interpretation of RQ2 Results}
The seven key quality attributes of AI-native applications identified in this study represent an effective extension of existing software quality models. The results of RQ2 indicate that software quality engineering is evolving from mere code verification toward a deep integration of data governance, model alignment, and probabilistic engineering.

The elevation of \textbf{Privacy and Safety} as an independent and high-priority theme reflects a fundamental change in the role of data in AI-native applications.  In traditional applications, privacy protection is usually regarded as a non-functional requirement or a compliance add-on.  In AI-native applications, data serve both as input and as the model’s memory.  The results show that developers are highly concerned about the risk of training data leakage and the privacy boundaries of context during inference [G4, G11, G20, G51, G53]. Additionally, this study emphasizes the critical role of \textbf{Observability}, highlighting the black-box dilemma in AI-native engineering.  Unlike traditional software, where errors can be traced through stack traces, failures in AI-native applications often stem from complex internal parameter interactions within the model and are inherently unexplainable. Therefore, high-quality AI-native applications must establish a new observability stack, including complete logging of prompts, replay of inference processes, and visualization of Chain-of-Thought (CoT) reasoning [G73]. Traditional software quality models~\cite{7848750} consider functional correctness as the core, meaning that a system should produce a deterministic and unique output for a given input.  However, for AI-native applications, \textbf{Reliability and Robustness} indicate that the core of quality has shifted toward managing uncertainty. Due to the probabilistic generation characteristics of LLMs, hallucinations have become an intrinsic feature rather than merely a code defect. Consequently, the focus of quality assurance is no longer on achieving zero defects, but on ensuring faithfulness of outputs and robustness under adversarial inputs. This implies that quality evaluation frameworks for AI-native applications should incorporate statistical metrics such as confidence intervals and consistency checks to quantify system reliability in non-deterministic environments. 

In addition, The close coupling of user experience and adaptability reveals the intent-centered nature of interaction in AI-native applications. High quality UX is no longer limited to interface layout. It depends on the system’s ability to understand vague natural language intentions and to proactively adapt to user preferences through memory and continuous learning. Performance and efficiency are no longer solely about speed and resource consumption for AI-native applications. They are intrinsically linked to economic viability.  The high cost of model inference makes cost-efficiency a primary architectural concern, creating a constant and complex trade-off between capability and operational expense.

\subsubsection{Interpretation of RQ3 Results}
The technical elements used in the development of
AI-native applications are not a simple list of tools but a coherent, multi-layered architecture. 

At the foundation layer, developers not only select \textbf{Core Model Technologies} but also reshape general intelligence into task-specific assets through \textbf{Model Training and Optimization}. They leverage \textbf{Local and Edge AI} to address privacy and latency concerns, while relying on \textbf{AI Infrastructure and Ops} to support large-scale training and inference. As noted in [G58], “Cloud: OpenAI… Local: Ollama” and “Optimized frameworks like TensorFlow Lite” [G1] illustrate the interchangeability of core models and the trend of multi-end adaptation. The application layer is responsible for transforming the probabilistic capabilities of underlying models into deterministic business value, forming the core of differentiation for AI-native applications. RAG provides the necessary long-term memory for AI-native applications through vector databases, while \textbf{Agentic Systems} endow applications with execution capabilities, including planning, tool use, and workflow automation. These two technical elements work in synergy to operationalize intelligence in concrete business scenarios. Data such as “Data Retrieval… Data Augmentation”[G34] and “LangGraph's agent-based task execution framework” [G53] show that developers are leveraging complex context management and task orchestration to constrain model hallucinations and uncertainty. At the interaction layer, \textbf{UI and Application Frameworks}, along with \textbf{Generative Media}, constitute the user-facing interface. The integration of multimodal capabilities enables the interface to be dynamically constructed based on user intent and model outputs, rather than being pre-rendered. These two technical elements collectively support a seamless, multimodal user experience.

While a sophisticated front end is crucial for delivering value, the true center of gravity in the current discourse is the application Layer. The high frequency of frameworks such as LangChain and techniques like RAG reveals that the core engineering challenge is not only the AI model itself but also the intelligent plumbing that connects the model to data, memory, and tools. With the frequent emergence of elements such as “agent; mcp” [G48] and “autonomous web research” [G53], the industry is accelerating from the passive chatbot paradigm toward proactive, agent-based intelligence. This occurs even as the current ecosystem remains in a critical transitional stage, moving from the fragmented competition of tools like Spring AI [G15] and LangChain [G53] toward interface standardization.   

Another key technical trend revealed by the results of RQ3 is the duality of deployment venues, cloud versus local.  The significant focus on local and on-device infrastructure reflects a direct response to the challenges of the cloud-centric model, including privacy, latency, and cost. This suggests that the future of AI-native applications is hybrid, with applications leveraging the most suitable environment for each task. Notably, the gap between observability quality attributes and the relatively low mention of specific tools further underscores a maturity gap, in which established best practices and standardized tooling have yet to emerge.

\subsubsection{Interpretation of RQ4 Results}
Synthesizing the opportunities and challenges reveals the central tension shaping the current AI-native applications landscape.  The opportunities are overwhelmingly technology driven and emerge from the bottom up. They center on what the technology enables, such as enhanced productivity, deep personalization, and innovation.  The overall narrative is one of immense potential, propelled by the capabilities of foundational models.

In contrast, the challenges are primarily pragmatic and arise from top down engineering and integration concerns. The highest ranked issues, performance, data privacy, and integration, reflect the enduring struggles of enterprise software development, now intensified by the complexities of AI. The result is a clear picture of a technology push colliding with the realities of embedding AI into existing and intricate business systems.

\subsection{Implications}
\subsubsection{Implications to practitioners}
The transition from traditional software to AI-native applications demands a fundamental shift in engineering mindset, from deterministic coding to probabilistic engineering. Practitioners must move beyond viewing AI as a mere feature add-on and instead treat it as a foundational architectural component.

\textbf{Capability-oriented system design.}
Traditional applications are typically designed and delivered around a set of predefined and relatively stable functions. In contrast, the 8 key characteristics of AI-native applications identified in this study, such as being AI-centric and continuously learning and adaptive, imply that system design should no longer aim at one-off functional delivery, but instead emphasize the ability to evolve continuously. Therefore, from the beginning of design, space needs to be reserved for ongoing prompt optimization and data-driven feedback loops. The system architecture should adopt a modular, observable, and rollback-enabled design to support rapid iteration and reliable evolution of models, prompts, and data throughout the entire lifecycle.

Decouple models and prompts from business logic to enable independent replacement and flexible composition. To address the non-determinism of generative models, introduce observability quality assurance mechanisms [G38], including robust logging for prompt or response pairs, automated evaluation pipelines for model drift and hallucination, and semantic metrics to ensure output quality and ethical safety throughout the application lifecycle. Support rapid rollback to previously stable versions when deploying new models or prompt strategies, so that risks can be controlled while enabling frequent experimentation and continuous optimization.

\textbf{Reliable system operation.} 
The 7 quality attributes of AI-native applications, such as privacy and adaptability, are no longer mere optional enhancements. They constitute essential prerequisites that directly determine whether a system is usable, deployable, and sustainably operable. Consequently, quality assurance mechanisms must be designed around these attributes. For instance, continuous evaluation of prompts can prevent performance degradation, human-in-the-loop mechanisms can be incorporated, and verification and intervention channels can be established for critical scenarios. Meanwhile, a purely cloud-centric strategy is increasingly insufficient for AI-native applications due to privacy, latency, and cost constraints. Practitioners should adopt a hybrid infrastructure mindset, leveraging scalable cloud resources for heavy reasoning tasks while utilizing local and on-device inference for privacy-sensitive or latency-critical operations [G1,G51]. This involves optimizing models for edge environments to balance performance with resource efficiency.

\textbf{Hybrid skill set.}
As generative AI technologies continue to evolve, developers’skill sets must expand accordingly. The core value of AI-native applications resides not only in foundational models but in the application layer that connects intelligence with proprietary data and user experiences. Practitioners should prioritize mastering orchestration frameworks, such as LangChain [G15], LlamaIndex [G91] and architectural patterns like RAG~\cite{ren2025investigating,arslan2024survey} and Agentic Workflows~\cite{li2024survey,talebirad2023multi} . The focus must shift from calling model APIs to designing the intelligent plumbing, , managing memory, leveraging tools, and optimizing context windows. Beyond traditional programming, practitioners need to acquire competencies in Prompt Engineering~\cite{liu2022design,white2023prompt,santu2023teler}, Fine-tuning~\cite{lin2024data,han2024parameterefficientfinetuninglargemodels}, and RLHF~\cite{zheng2023secrets,ziegler2019fine}. Success lies in discerning when to apply specific techniques, such as using RAG for dynamic knowledge retrieval, fine-tuning for domain-specific style adaptation, or agentic frameworks for complex task planning.

\subsubsection{Implications to researchers}
The distinction between AI-enabled applications and AI-native applications marks a critical frontier for research. Scholars should pivot from "AI for SE" to "SE for AI," addressing the unique challenges posed by this new paradigm.

\textbf{AI-native software engineering.}
Hassan et al. proposed AI-native software engineering (SE 3.0), characterized by an intent-centered, conversation-driven development process that emphasizes collaboration between human developers and AI teammates~\cite{hassan2024towards}. However, the industry has yet to reach a consensus on a unified definition of AI-native applications, and academic research on this domain remains scarce. Significant research focuses on using generative AI to optimize software maintenance, such as code review~\cite{yu2024fine}, architecture detection~\cite{cao2025enhancing}, etc. There is still a gap in understanding the engineering of systems where the AI model is the core logic engine. Future research could clarify the methodological and practical differences among traditional software engineering (SE 1.0), AI-enabled software engineering (SE 2.0), and AI-native software engineering (SE 3.0) by conducting comparative analyses of their architectural patterns, lifecycle evolution, and failure modes. Moreover, drawing on the historical evolution of cloud-native applications can provide a theoretical perspective for understanding the developmental stages of AI-native applications, treating them as a distinct phase in the evolution of information systems, and facilitating the identification of technical, architectural, and organizational challenges.

\textbf{AI-native applications architecture.}
AI-native applications are not only concerned with single-turn task completion, but place greater emphasis on multi-turn reasoning capabilities, the effectiveness of tool use, and the semantic relevance of retrieved context. They also face unique challenges such as model hallucinations, inconsistent interactions, and long-term contextual dependencies. Therefore, there is an urgent need to develop systematic design and evaluation methods. First, at the architectural level, it is necessary to explore new architectural patterns suitable for AI-native applications, while also identifying potential anti-patterns to avoid design pitfalls. In addition, attention should be paid to the system’s ability to evolve over long-term iterations, including model updates, knowledge base expansion, and optimization of multi-turn interaction strategies. This not only helps establish a systematic approach to AI-native architectural design, but also provides guidance for subsequent quality evaluation and performance optimization. As the industry adopts complex agentic and RAG architectures [G41, G55, G65], traditional accuracy metrics, such as F1 score, are no longer sufficient to comprehensively measure system performance. Therefore, there is a need to establish standardized benchmarks for quantifying agentic success and methodologies for stress-testing autonomous workflows against unintended behaviors. In addition, the quality attributes of AI-native applications should be formally defined, and based on these definitions, automated evaluation methods and benchmark datasets should be developed to provide a scientific foundation for system quality assessment.

\textbf{AI-native software development lifecycle.}
In existing academic research, only a very small number of studies have systematically discussed the software lifecycle of AI-native applications. Building on the traditional V-model, Hymel~\cite{hymel2024ai} proposed the V-Bounce model, which seamlessly integrates AI into every development phase from planning to deployment, and shifts the focus from one-time acceptance to requirements elicitation, architectural design, and continuous validation. This reflects a shift in the development of AI-native applications from a linear process to a continuously evolving process. Correspondingly, the results of this study indicate that AI-native applications exhibit a dynamic evolutionary paradigm that continuously adapts to changes in data, models, and environments. Therefore, it is necessary to further explore development process models tailored for AI-native applications to support continuous evaluation and online evolution. Current MLOps practices~\cite{kreuzberger2023machine} are often model-centric rather than system-centric, making it difficult to support such continuously evolving application forms. Future research needs to expand toward LLMOps or AgentOps, treating AI-native applications as complex socio-technical systems, and systematically studying the interactions over time among application code, orchestration logic, and probabilistic models. This would enable the development of methodologies to ensure long-term system stability in the face of foundation model updates and data distribution shifts.

\subsection{Threats to Validity}
\label{sec6}

Following the guidelines for validity analysis \cite{shull2008guide,wohlin2012experimentation}, this section evaluates potential threats to the validity of the research findings.

\textbf{Internal Validity:} the primary threat stems from the inherently interpretive nature of coding and abstracting data into meaningful themes. A researcher's pre-existing beliefs or expectations could unconsciously influence which data segments are considered significant or how themes are categorized and named. To address this threat, we followed the structured, multi-stage methodology proposed by Braun and Clarke~\cite{cruzes2011recommended}. This systematic process, extending from initial data familiarization to the final definition of themes, provided a consistent framework that guided the analysis and limited purely subjective interpretations. In addition, we maintained a high level of transparency in our reporting. By presenting detailed evidence of the analytical process, including tables of initial codes, candidate themes, and final themes with their corresponding frequency counts, we created a clear audit trail. This transparency enables readers to trace the evidence from raw data segments to high-level conclusions, thereby enhancing the trustworthiness and internal validity of the findings.

\textbf{External Validity:} the study relies on publicly accessible grey literature, which may introduce selection bias. The focus on English-language sources that are prominent in search engines or open-source communities could underrepresent non-English perspectives, unpublished academic research, or organizations that do not actively publish thought-leadership content. Grey literature, particularly corporate blogs and venture capital materials, often emphasizes opportunities while downplaying long-term maintenance challenges or project failures. The AI-native field is highly dynamic, and this study captures only a snapshot of discourse within a specific timeframe. The relevance of technologies, the prominence of quality attributes, and the nature of the challenges may change rapidly. To address these limitations, we deliberately sampled a diverse set of sources including technology companies, independent practitioners, open-source project documentation, and venture capital perspectives, allowing us to triangulate viewpoints and balance aspirational visions with concrete technical issues. We also incorporated challenges identified in project issues, which provided a critical counterbalance to the optimism of blogs. Finally, we have been explicit that the findings reflect the dominant discourse within a defined scope, and by presenting the study as a timely analysis of the current paradigm, we acknowledge its temporal limitations while underscoring its contemporary relevance.

\textbf{Construct Validity:} as a nascent and rapidly evolving concept, AI-Native does not yet have a universally accepted or formal definition.  Different authors employ the term with subtle variations in meaning, which can lead to the conflation of related but distinct ideas.  To mitigate this, we deliberately refrained from imposing a rigid a priori definition and instead employed Braun and Clarke's thematic analysis method \cite{cruzes2011recommended}, which allows the definition and its core components to emerge inductively from the data.  By synthesizing definitions and characteristics across a wide range of sources, this study constructs a bottom-up, evidence-based account of how the term is currently understood and applied in practice.

\textbf{Conclusion validity:} the primary threat in this qualitative context is the potential for misinterpreting thematic patterns or overstating the certainty of our findings. To address this, we employed several strategies. We transparently used frequency counts not as statistical proof, but as a robust indicator of a theme's prominence within the discourse, providing a clear and evidence-based rationale for our analytical focus. In addition, our most significant conclusions are supported through triangulation, drawing insights not from individual themes in isolation, but from consistent patterns observed across multiple, interrelated areas of the analysis.
\section{Conclusions and Future Work}
\label{sec7}
This study maps the emergent AI-native applications paradigm, presenting it as a transformative evolution in software engineering defined by intelligent automation, continuous learning, and multimodal capabilities.  AI-native applications fundamentally reshape software quality by shifting attention from deterministic code to complex, probabilistic, and costly systems where observability, reliability, and economic efficiency are paramount.  However, immense technology-driven opportunities for innovation and productivity are constrained by integration challenges, stability concerns, and a gap in articulating measurable business value.  AI-native applications lie at the intersection of technological possibility and operational practicality and represent a highly promising yet still maturing domain.

Several important directions emerge from these findings. AI-native observability requires standardized frameworks and tools capable of monitoring agentic workflows, RAG pipelines, and model outputs in real time. Research is also needed on the migration from Cloud-Native to AI-Native architectures, including the adaptation of orchestration layers, service meshes, and APIs to manage probabilistic services, guided by longitudinal case studies. Post-deployment monitoring and long-term maintenance constitute another critical area, demanding methods for detecting concept drift, managing dependencies among code, data, prompts, and models, and evaluating the economics of maintaining these evolving systems.

\bibliographystyle{cas-model2-names}
\bibliography{cas-refs}

\clearpage
\appendix
\renewcommand{\thesection}{\Alph{section}}
\counterwithin{table}{section}
\section{Selected GL Sources}
\noindent
[G1]Sirota, A., \textit{What are AI-native mobile apps, and what are their use cases?}, Pluralsight, 2025.
\url{https://www.pluralsight.com/resources/blog/ai-and-data/ai-native-mobile-apps}.

\vspace{0.5em}
\noindent
[G2]Dharani, N., Kim, Y., Choudhari, A., Ron, S.-L., Bhatia, A., Misra, N., \textit{From Apps to Agents: How the AI-Native Tech Stack Is Transforming Software}, medium, 2025.
\url{https://medium.com/bcgontech/from-apps-to-agents-how-the-ai-native-tech-stack-is-transforming-software-c34ab557d228}.

\vspace{0.5em}
\noindent
[G3]Shenoy, N., \textit{AI Native — What Does It Mean for Embedded Processing?}, synaptics, 2025.
\url{https://www.synaptics.com/company/blog/ai-native-embedded-processing}.

\vspace{0.5em} 
\noindent
[G4]Shimel, A., \textit{AI Native Dev: Shaping the Future of AI-First Software Development}, devops, 2025.
\url{https://devops.com/ai-native-dev-shaping-the-future-of-ai-first-software-development/}.

\vspace{0.5em} 
\noindent
[G5]r/msp, \textit{What AI native stack replacement companies are on your radar?}, reddit, 2025.
\url{https://www.reddit.com/r/msp/comments/1k3nyxs/what_ai_native_stack_replacement_companies_are_on/}.

\vspace{0.5em} 
\noindent
[G6]Hypermode, \textit{What does AI-native mean?}, Hypermode, 2025.
\url{https://hypermode.com/blog/ai-native-guide}.

\vspace{0.5em}
\noindent
[G7]Abnormal, \textit{What Is AI Native? Designing AI-Powered Cybersecurity Systems}, abnormal, 2025.
\url{https://abnormal.ai/ai-glossary/ai-native}.

\vspace{0.5em}
\noindent
[G8]Smeyatsky, A., \textit{AI-Native Architecture: Definition, Core Concepts, and Comparison with Cloud Native}, linkedin, 2025.
\url{https://www.linkedin.com/pulse/ai-native-architecture-definition-core-concepts-cloud-allan-smeyatsky-qgamf}.

\vspace{0.5em}
\noindent
[G9]Swimm Team, \textit{What Is AI Native? Benefits, Use Cases, and Best Practices}, swimm, 2025.
\url{https://swimm.io/learn/software-development/what-is-ai-native-benefits-use-cases-and-best-practices}.

\vspace{0.5em}
\noindent
[G10]Hypermode, \textit{How to get started with AI-native application development}, hypermode, 2025.
\url{https://hypermode.com/blog/ai-native-app-development-guide}.

\vspace{0.5em}
\noindent
[G11]Riaz, H., \textit{AI-Native vs. AI-Included: What’s the Difference and Why It Matters for Sustainability}, ensogo, 2025.
\url{https://ensogo.io/ai-native-vs-ai-included-whats-the-difference-and-why-it-matters-for-sustainability/}.

\vspace{0.5em}
\noindent
[G12]Budnik, O., \textit{The rise of AI-native technologies}, altamira, 2025.
\url{https://www.altamira.ai/blog/rise-of-ai-native-technologies/}.

\vspace{0.5em}
\noindent
[G13]Ferrer, Ó., \textit{AI Native Applications}, paradigmadigital, 2025.
\url{https://en.paradigmadigital.com/techbiz/ai-native-applications/}.

\vspace{0.5em}
\noindent
[G14]Rapal, H., \textit{What Is an AI-Native CRM and Why Your Business Needs One}, legittai, 2025.
\url{https://legittai.com/blog/what-is-an-ai-native-crm}.

\vspace{0.5em}
\noindent
[G15]Kumar, A., \textit{What is AI-Native Software Engineering and Why Should You Care?}, medium, 2025.
\url{https://codefarm0.medium.com/what-is-ai-native-software-engineering-and-why-should-you-care-f04f268c8f7a}.

\vspace{0.5em}
\noindent
[G16]Editorial Staff, \textit{AI Native in Action: Why It’s the Key to Future Technological Success}, getgenie, 2025.
\url{https://getgenie.ai/ai-native-in-action/}.

\vspace{0.5em}
\noindent
[G17]Mindverse, \textit{Second-Me}, GitHub, 2025.
\url{https://github.com/mindverse/Second-Me}.

\vspace{0.5em}
\noindent
[G18]nickscamara, \textit{open-deep-research}, GitHub, 2025.
\url{https://github.com/nickscamara/open-deep-research}.

\vspace{0.5em}
\noindent
[G19]GoogleCloudPlatform, \textit{kubectl-ai}, GitHub, 2025.
\url{https://github.com/GoogleCloudPlatform/kubectl-ai}.

\vspace{0.5em}
\noindent
[G20]Mirix-AI, \textit{MIRIX}, GitHub, 2025.
\url{https://github.com/Mirix-AI/MIRIX}.

\vspace{0.5em}
\noindent
[G21]AnotiaWang, \textit{deep-research-web-ui}, GitHub, 2025.
\url{https://github.com/AnotiaWang/deep-research-web-ui}.

\vspace{0.5em}
\noindent
[G22]Vexa-ai, \textit{vexa}, GitHub, 2025.
\url{https://github.com/Vexa-ai/vexa}.

\vspace{0.5em}
\noindent
[G23]rikkahub, \textit{rikkahub}, GitHub, 2025.
\url{https://github.com/rikkahub/rikkahub}.

\vspace{0.5em}
\noindent
[G24]sagekit, \textit{magnitude}, GitHub, 2025.
\url{https://github.com/magnitudedev/magnitude}.

\vspace{0.5em}
\noindent
[G25]Stability-AI, \textit{stable-virtual-camera}, GitHub, 2025.
\url{https://github.com/Stability-AI/stable-virtual-camera}.

\vspace{0.5em}
\noindent
[G26]bytebot-ai, \textit{bytebot}, GitHub, 2025.
\url{https://github.com/bytebot-ai/bytebot}.

\vspace{0.5em}
\noindent
[G27]neuphonic, \textit{neutts-air}, GitHub, 2025.
\url{https://github.com/neuphonic/neutts-air}.

\vspace{0.5em}
\noindent
[G28]bytedance, \textit{InfiniteYou}, GitHub, 2025.
\url{https://github.com/bytedance/InfiniteYou}.

\vspace{0.5em}
\noindent
[G29]Phantom-video, \textit{Phantom}, GitHub, 2025.
\url{https://github.com/Phantom-video/Phantom}.

\vspace{0.5em}
\noindent
[G30]Ism, I., \textit{The Rise of AI-Native Applications}, Voicenotes, 2024.
\url{https://voicenotes.com/blog/ai-native}.

\vspace{0.5em}
\noindent
[G31]Gao, C., Burke, K., Segall, L., DellaPasqua, J., Reddy, A., Liao, M., \textit{AI-Native Applications: A Framework for Evaluating the Future of Enterprise Software}, Sapphire Ventures, 2024.
\url{https://sapphireventures.com/blog/ai-native-applications/}.

\vspace{0.5em}
\noindent
[G32]Zelnikov, A., \textit{Rethinking application architecture: my vision of AI-native apps}, LinkedIn, 2024.
\url{https://www.linkedin.com/pulse/rethinking-application-architecture-my-vision-apps-andrei-zelnikov-mzczc}.

\vspace{0.5em}
\noindent
[G33]Meabe, C., \textit{AI-Native vs. Embedded AI: Unraveling the Core Differences}, Foundationinc, 2024.
\url{https://foundationinc.co/lab/ai-native-embedded-ai}.

\vspace{0.5em}
\noindent
[G34]Meabe, C., \textit{What the F**k is an AI-native app?}, Unbody, 2024.
\url{https://unbody.io/blog/ai-native-app}.

\vspace{0.5em}
\noindent
[G35]Hanxie, \textit{AI-native Applications Based on Event-driven Architectures}, Alibaba Cloud, 2024.
\url{https://www.alibabacloud.com/blog/ai-native-applications-based-on-event-driven-architectures_601838}.

\vspace{0.5em}
\noindent
[G36]Bohdankit, \textit{WTF is AI-Native Product?}, Bohdan Kit, 2024.
\url{https://bohdankit.com/wtf-is-ai-native-product/}.

\vspace{0.5em}
\noindent
[G37]IrisAgent, \textit{AI Native: The Future of Enterprise Innovation}, IrisAgent, 2024.
\url{https://irisagent.com/blog/ai-native-the-future-of-enterprise-innovation/}.

\vspace{0.5em}
\noindent
[G38]Alibaba Cloud Native, \textit{New Scenarios and New Capabilities: Observability Innovations in the AI-Native Era}, Alibaba Cloud, 2024.
\url{https://www.alibabacloud.com/blog/new-scenarios-and-new-capabilities-observability-innovations-in-the-ai-native-era_601687}.

\vspace{0.5em}
\noindent
[G39]Raza, M., \textit{What Is AI Native?}, splunk, 2024.
\url{https://www.splunk.com/en_us/blog/learn/ai-native.html}.

\vspace{0.5em}
\noindent
[G40]Nucci, A., \textit{AI Native Explained}, aisera, 2024.
\url{https://aisera.com/blog/ai-native/}.

\vspace{0.5em}
\noindent
[G41]Borthwick, J., \textit{Announcing AI Camp: Native Applications}, medium, 2024.
\url{https://render.betaworks.com/announcing-ai-camp-native-applications-e1358061c601}.

\vspace{0.5em}
\noindent
[G42]McQueen, R., \textit{Understanding AI-Native Applications: Technical Analysis and Real-World Impact}, openkit, 2024.
\url{https://openkit.ai/posts/technical-analysis-ai-native-applications}.

\vspace{0.5em}
\noindent
[G43]Gallagher, C., \textit{Unleashing the Power of AI: The Evolution of AI Native Data Infrastructure}, weka, 2024.
\url{https://www.weka.io/blog/ai-ml/unleashing-the-power-of-ai-the-evolution-of-ai-native-data-infrastructure/}.

\vspace{0.5em}
\noindent
[G44]AlShikh, W., \textit{Why AI-native enterprise apps are the business brain of the future}, writer, 2024.
\url{https://writer.com/engineering/ai-native-apps/}.

\vspace{0.5em}
\noindent
[G45]Cruise, D., \textit{Navigating the future of enterprise systems: Cloud Native vs AI Native ERP}, changeassociates, 2024.
\url{https://changeassociates.com/cloud-native-vs-ai-native-erp/}.

\vspace{0.5em}
\noindent
[G46]Hymel, C., \textit{The AI-Native Software Development Lifecycle: A Theoretical and Practical New Methodology}, arXiv, 2024.
\url{https://arxiv.org/abs/2408.03416}.

\vspace{0.5em}
\noindent
[G47]Hassan, A. E., Oliva, G. A., Lin, D., Chen, B., Jiang, Z. M., \textit{Towards AI-Native Software Engineering (SE 3.0): A Vision and a Challenge Roadmap}, arXiv, 2024.
\url{https://arxiv.org/abs/2410.06107}.

\vspace{0.5em}
\noindent
[G48]codexu, \textit{note-gen}, GitHub, 2024.
\url{https://github.com/codexu/note-gen}.

\vspace{0.5em}
\noindent
[G49]Zackriya-Solutions, \textit{meeting-minutes}, GitHub, 2024.
\url{https://github.com/Zackriya-Solutions/meeting-minutes}.

\vspace{0.5em}
\noindent
[G50]languine-ai, \textit{languine}, GitHub, 2024.
\url{https://github.com/languine-ai/languine}.

\vspace{0.5em}
\noindent
[G51]a-ghorbani, \textit{pocketpal-ai}, GitHub, 2024.
\url{https://github.com/a-ghorbani/pocketpal-ai}.

\vspace{0.5em}
\noindent
[G52]refly-ai, \textit{refly}, GitHub, 2024.
\url{https://github.com/refly-ai/refly}.

\vspace{0.5em}
\noindent
[G53]ai-christianson, \textit{RA.Aid}, GitHub, 2024.
\url{https://github.com/ai-christianson/RA.Aid}.

\vspace{0.5em}
\noindent
[G54]fal-ai-community, \textit{video-starter-kit}, GitHub, 2024.
\url{https://github.com/fal-ai-community/video-starter-kit}.

\vspace{0.5em}
\noindent
[G55]nanbingxyz, \textit{5ire}, GitHub, 2024.
\url{https://github.com/nanbingxyz/5ire}.

\vspace{0.5em}
\noindent
[G56]moeru-ai, \textit{airi}, GitHub, 2024.
\url{https://github.com/moeru-ai/airi}.

\vspace{0.5em}
\noindent
[G57]yetone, \textit{avante.nvim}, GitHub, 2024.
\url{https://github.com/yetone/avante.nvim}.

\vspace{0.5em}
\noindent
[G58]theJayTea, \textit{WritingTools}, GitHub, 2024.
\url{https://github.com/theJayTea/WritingTools}.

\vspace{0.5em}
\noindent
[G59]Doriandarko, \textit{deepseek-engineer}, GitHub, 2024.
\url{https://github.com/Doriandarko/deepseek-engineer}.

\vspace{0.5em}
\noindent
[G60]raidendotai, \textit{cofounder}, GitHub, 2024.
\url{https://github.com/raidendotai/cofounder}.

\vspace{0.5em}
\noindent
[G61]sentient-agi, \textit{OML-1.0-Fingerprinting}, GitHub, 2024.
\url{https://github.com/sentient-agi/OML-1.0-Fingerprinting}.

\vspace{0.5em}
\noindent
[G62]johnbean393, \textit{Sidekick}, GitHub, 2024.
\url{https://github.com/johnbean393/Sidekick}.

\vspace{0.5em}
\noindent
[G63]Beingpax, \textit{VoiceInk}, GitHub, 2024.
\url{https://github.com/Beingpax/VoiceInk}.

\vspace{0.5em}
\noindent
[G64]wandb, \textit{openui}, GitHub, 2024.
\url{https://github.com/wandb/openui}.

\vspace{0.5em}
\noindent
[G65]X-PLUG, \textit{MobileAgent}, GitHub, 2024.
\url{https://github.com/X-PLUG/MobileAgent}.

\vspace{0.5em}
\noindent
[G66]emcf, \textit{thepipe}, GitHub, 2024.
\url{https://github.com/emcf/thepipe}.

\vspace{0.5em}
\noindent
[G67]nottelabs, \textit{notte}, GitHub, 2024.
\url{https://github.com/nottelabs/notte}.

\vspace{0.5em}
\noindent
[G68]itsOwen, \textit{CyberScraper-2077}, GitHub, 2024.
\url{https://github.com/itsOwen/CyberScraper-2077}.

\vspace{0.5em}
\noindent
[G69]ToTheBeginning, \textit{PuLID}, GitHub, 2024.
\url{https://github.com/ToTheBeginning/PuLID}.

\vspace{0.5em}
\noindent
[G70]adrianhajdin, \textit{ai\_saas\_app}, GitHub, 2024.
\url{https://github.com/adrianhajdin/ai_saas_app}.

\vspace{0.5em}
\noindent
[G71]xming521, \textit{WeClone}, GitHub, 2024.
\url{https://github.com/xming521/WeClone}.

\vspace{0.5em}
\noindent
[G72]swark-io, \textit{swark}, GitHub, 2024.
\url{https://github.com/swark-io/swark}.

\vspace{0.5em}
\noindent
[G73]AI4Finance-Foundation, \textit{FinRobot}, GitHub, 2024.
\url{https://github.com/AI4Finance-Foundation/FinRobot}.

\vspace{0.5em}
\noindent
[G74]huggingface, \textit{speech-to-speech}, GitHub, 2024.
\url{https://github.com/huggingface/speech-to-speech}.

\vspace{0.5em}
\noindent
[G75]Skyvern-AI, \textit{skyvern}, GitHub, 2024.
\url{https://github.com/Skyvern-AI/skyvern}.

\vspace{0.5em}
\noindent
[G76]alibaba, \textit{Tora}, GitHub, 2024.
\url{https://github.com/alibaba/Tora}.

\vspace{0.5em}
\noindent
[G77]TMElyralab, \textit{MuseV}, GitHub, 2024.
\url{https://github.com/TMElyralab/MuseV}.

\vspace{0.5em}
\noindent
[G78]1038lab, \textit{ComfyUI-RMBG}, GitHub, 2024.
\url{https://github.com/1038lab/ComfyUI-RMBG}.

\vspace{0.5em}
\noindent
[G79]Iovene, M., Jonsson, L., Roeland, D., D’Angelo, M., \textit{Defining AI native: A key enabler for advanced intelligent telecom networks}, Ericsson, 2023.
\url{https://www.ericsson.com/en/reports-and-papers/white-papers/ai-native}.

\vspace{0.5em}
\noindent
[G80]DeepInsight-AI, \textit{DeepBI}, GitHub, 2023.
\url{https://github.com/DeepInsight-AI/DeepBI}.

\vspace{0.5em}
\noindent
[G81]Mintplex-Labs, \textit{anything-llm}, GitHub, 2023.
\url{https://github.com/Mintplex-Labs/anything-llm}.

\vspace{0.5em}
\noindent
[G82]Navigation Menu, \textit{generative-ai-use-cases}, GitHub, 2023.
\url{https://github.com/aws-samples/generative-ai-use-cases}.

\vspace{0.5em}
\noindent
[G83]jupyterlab, \textit{jupyter-ai}, GitHub, 2023.
\url{https://github.com/jupyterlab/jupyter-ai}.

\vspace{0.5em}
\noindent
[G84]Coframe, \textit{coffee}, GitHub, 2023.
\url{https://github.com/Coframe/coffee}.

\vspace{0.5em}
\noindent
[G85]myshell-ai, \textit{OpenVoice}, GitHub, 2023.
\url{https://github.com/myshell-ai/OpenVoice}.

\vspace{0.5em}
\noindent
[G86]lllyasviel, \textit{Fooocus}, GitHub, 2023.
\url{https://github.com/lllyasviel/Fooocus}.

\vspace{0.5em}
\noindent
[G87]aress31, \textit{burpgpt}, GitHub, 2023.
\url{https://github.com/aress31/burpgpt}.

\vspace{0.5em}
\noindent
[G88]openai, \textit{chatgpt-retrieval-plugin}, GitHub, 2023.
\url{https://github.com/openai/chatgpt-retrieval-plugin}.

\vspace{0.5em}
\noindent
[G89]h2oai, \textit{h2ogpt}, GitHub, 2023.
\url{https://github.com/h2oai/h2ogpt}.

\vspace{0.5em}
\noindent
[G90]collabora, \textit{WhisperLive}, GitHub, 2023.
\url{https://github.com/collabora/WhisperLive}.

\vspace{0.5em}
\noindent
[G91]szczyglis-dev, \textit{py-gpt}, GitHub, 2023.
\url{https://github.com/szczyglis-dev/py-gpt}.

\vspace{0.5em}
\noindent
[G92]TencentARC, \textit{PhotoMaker}, GitHub, 2023.
\url{https://github.com/TencentARC/PhotoMaker}.

\vspace{0.5em}
\noindent
[G93]logancyang, \textit{obsidian-copilot}, GitHub, 2023.
\url{https://github.com/logancyang/obsidian-copilot}.

\vspace{0.5em}
\noindent
[G94]olimorris, \textit{codecompanion.nvim}, GitHub, 2023.
\url{https://github.com/olimorris/codecompanion.nvim}.

\vspace{0.5em}
\noindent
[G95]keephq, \textit{keep}, GitHub, 2023.
\url{https://github.com/keephq/keep}.

\vspace{0.5em}
\noindent
[G96]Doubiiu, \textit{DynamiCrafter}, GitHub, 2023.
\url{https://github.com/Doubiiu/DynamiCrafter}.

\vspace{0.5em}
\noindent
[G97]plasma-umass, \textit{ChatDBG}, GitHub, 2023.
\url{https://github.com/plasma-umass/ChatDBG}.

\vspace{0.5em}
\noindent
[G98]unum-cloud, \textit{Uform}, GitHub, 2023.
\url{https://github.com/unum-cloud/uform}.

\vspace{0.5em}
\noindent
[G99]acon96, \textit{home-llm}, GitHub, 2023.
\url{https://github.com/acon96/home-llm}.

\vspace{0.5em}
\noindent
[G100]instantX-research, \textit{InstantStyle}, GitHub, 2023.
\url{https://github.com/instantX-research/InstantStyle}.

\vspace{0.5em}
\noindent
[G101]InternLM, \textit{InternLM-Xcomposer}, GitHub, 2023.
\url{https://github.com/InternLM/InternLM-Xcomposer}.

\vspace{0.5em}
\noindent
[G102]Future-House, \textit{paper-qa}, GitHub, 2023.
\url{https://github.com/Future-House/paper-qa}.

\vspace{0.5em}
\noindent
[G103]eyurtsev, \textit{kor}, GitHub, 2023.
\url{https://github.com/eyurtsev/kor}.

\vspace{0.5em}
\noindent
[G104]phodal, \textit{auto-dev}, GitHub, 2023.
\url{https://github.com/phodal/auto-dev}.

\vspace{0.5em}
\noindent
[G105]activepieces, \textit{activepieces}, GitHub, 2022.
\url{https://github.com/activepieces/activepieces}.

\vspace{0.5em}
\noindent
[G106]vladmandic, \textit{sdnext}, GitHub, 2022.
\url{https://github.com/vladmandic/sdnext}.

\section{Quality Assessment Criteria for GL}
\label{app1}
\clearpage
\begin{table*}[!htbp]
\centering
\caption{Quality assessment criteria}
\label{taba}
\begin{tabularx}{\textwidth}{l|>{\raggedright\arraybackslash}X}
\hline
\textbf{Criteria} & \textbf{Questions} \\
\hline
\multirow{4}{*}{Authority of the producer} 
 & Is the publishing organization reputable? For example, the Software Engineering Institute (SEI) \\ \cline{2-2} 
 & Is an individual author associated with a reputable organization? \\ \cline{2-2} 
 & Has the author published other work in the field? \\ \cline{2-2} 
 & Does the author have expertise in the area? (e.g., job title principal software engineer) \\ \hline

\multirow{6}{*}{Methodology} 
 & Does the source have a clearly stated aim? \\ \cline{2-2} 
 & Does the source have a stated methodology? \\ \cline{2-2} 
 & Is the source supported by authoritative, contemporary references? \\ \cline{2-2} 
 & Are any limits clearly stated? \\ \cline{2-2} 
 & Does the work cover a specific question? \\ \cline{2-2} 
 & Does the work refer to a particular population or case? \\ \hline

\multirow{4}{*}{Objectivity} 
 & Does the work seem to be balanced in presentation? \\ \cline{2-2} 
 & Is the statement in the sources as objective as possible? Or, is the statement a subjective opinion? \\ \cline{2-2} 
 & Is there a vested interest? For example, a tool comparison by authors working for a particular tool vendor \\ \cline{2-2} 
 & Are the conclusions supported by the data? \\ \hline

Date & Does the item have a clearly stated date? \\ \hline

Position w.r.t. related sources & Have key-related GL or formal sources been linked to/discussed? \\ \hline

\multirow{2}{*}{Novelty} 
 & Does it enrich or add something unique to the research? \\ \cline{2-2} 
 & Does it strengthen or refute a current position? \\ \hline

Impact & Normalize all the following impact metrics into a single aggregated impact metric (when data are available): number of citations; number of backlinks; number of social media shares (alt-metrics); number of comments posted for a specific online entry, like a blog post or a video; number of page or paper views \\ \hline

\multirow{3}{*}{Outlet type} 
 & First tier GL (measure = 1): high outlet control/high credibility: books, magazines, theses, government reports, white papers \\ \cline{2-2} 
 & Second tier GL (measure = 0.5): moderate outlet control/moderate credibility: annual reports, news articles, presentations, videos, Q/A sites (e.g., Stack Overflow), Wiki articles \\ \cline{2-2} 
 & Third tier GL (measure = 0): low outlet control/low credibility: blogs, emails, tweets \\ \hline
\end{tabularx}
\end{table*}


\end{document}